\documentstyle[12pt]{article}
\begin{document}


\def\Nequalstwo{\Psi}
\def\eff{{\rm eff}}
\def\inst{{\rm inst}}
\def\fermi{{\rm fermi}}
\def\trtwo{\tr^{}_2\,}
\def\finv{f^{-1}}
\def\Ubar{\bar U}
\def\wbar{\bar w}
\def\fbar{\bar f}
\def\abar{\bar a}
\def\bbar{\bar b}
\def\Deltabar{\bar\Delta}
\def\dalpha{{\dot\alpha}}
\def\dbeta{{\dot\beta}}
\def\dgamma{{\dot\gamma}}
\def\ddelta{{\dot\delta}}
\def\Sbar{\bar S}
\def\Im{{\rm Im}}
\def\sst{\scriptscriptstyle}
\def\cld{C_{\sst\rm LD}^{}}
\def\csd{C_{\sst\rm SD}^{}}
\def\bigI{{\rm I}_{\sst 3\rm D}}
\def\Mr{{\rm M}_{\sst R}}
\def\cJ{C_{\sst J}}
\def\one{{\sst(1)}}
\def\two{{\sst(2)}}
\def\vsd{v^{\sst\rm SD}}
\def\vasd{v^{\sst\rm ASD}}
\def\Phibar{\bar\Phi}
\def\F{{\cal F}_{\sst\rm SW}}
\def\P{{\cal P}}
\def\A{{\cal A}}
\def\susy{supersymmetry}
\def\sigmabar{\bar\sigma}
\def\barsigma{\sigmabar}
\def\ASD{{\scriptscriptstyle\rm ASD}}
\def\cl{{\,\rm cl}}
\def\lambdabar{\bar\lambda}
\def\R{{R}}
\def\psibar{\bar\psi}
\def\sqrtwo{\sqrt{2}\,}
\def\etabar{\bar\eta}
\def\Thetabar{{\bar\Theta_0}}
\def\Qbar{\bar Q}
\def\susic{supersymmetric}
\def\vhiggs{{\rm v}}
\def\vhiggsa{{\cal A}_{\sst00}}
\def\vbarhiggs{\bar{\rm v}}
\def\vhiggsbar{\bar{\rm v}}
\def\novetal{Novikov et al.}
\def\Novetal{Novikov et al.}
\def\ADS{Affleck, Dine and Seiberg}
\def\ads{Affleck, Dine and Seiberg}
\def\setI{\{{\cal I}\}}
\def\Abar{A^\dagger}
\def\B{{\cal B}}
\def\infinity{\infty}
\def\C{{\cal C}}
\def\Psitwo{\Psi_{\scriptscriptstyle N=2}}
\def\Psibartwo{\bar\Psi_{\scriptscriptstyle N=2}}
\def\zero{{\scriptscriptstyle(0)}}
\def\new{{\scriptscriptstyle\rm new}}
\def\u{\underline}
\def\uA{\,\lower 1.2ex\hbox{$\sim$}\mkern-13.5mu A}
\def\uBmu{\,\lower 1.2ex\hbox{$\sim$}\mkern-13.5mu B_\mu}
\def\uAmu{\,\lower 1.2ex\hbox{$\sim$}\mkern-13.5mu A_\mu}
\def\uX{\,\lower 1.2ex\hbox{$\sim$}\mkern-13.5mu X}
\def\uD{\,\lower 1.2ex\hbox{$\sim$}\mkern-13.5mu {\rm D}}
\def\uDzero{{\uD}^\zero}
\def\uAzero{{\uA}^\zero}
\def\upsizero{{\upsi}^\zero}
\def\uF{\,\lower 1.2ex\hbox{$\sim$}\mkern-13.5mu F}
\def\uW{\,\lower 1.2ex\hbox{$\sim$}\mkern-13.5mu W}
\def\uWbar{\,\lower 1.2ex\hbox{$\sim$}\mkern-13.5mu {\overline W}}
\def\Dbar{D^\dagger}
\def\Fbar{F^\dagger}
\def\uAbar{{\uA}^\dagger}
\def\uAbarzero{{\uA}^{\dagger\zero}}
\def\uDbar{{\uD}^\dagger}
\def\uDbarzero{{\uD}^{\dagger\zero}}
\def\uFbar{{\uF}^\dagger}
\def\uFbarzero{{\uF}^{\dagger\zero}}
\def\uV{\,\lower 1.2ex\hbox{$\sim$}\mkern-13.5mu V}
\def\uZ{\,\lower 1.2ex\hbox{$\sim$}\mkern-13.5mu Z}
\def\uv{\lower 1.0ex\hbox{$\scriptstyle\sim$}\mkern-11.0mu v}
\def\uc{\lower 1.0ex\hbox{$\scriptstyle\sim$}\mkern-11.0mu c}
\def\uB{\lower 1.0ex\hbox{$\scriptstyle\sim$}\mkern-11.0mu B}
\def\uPsi{\,\lower 1.2ex\hbox{$\sim$}\mkern-13.5mu \Psi}
\def\uPhi{\,\lower 1.2ex\hbox{$\sim$}\mkern-13.5mu \Phi}
\def\uchi{\lower 1.5ex\hbox{$\sim$}\mkern-13.5mu \chi}
\def\utheta{\lower 1.5ex\hbox{$\sim$}\mkern-13.5mu \theta}
\def\chitilde{\tilde \chi}
\def\etatilde{\tilde \eta}
\def\uchitilde{\lower 1.5ex\hbox{$\sim$}\mkern-13.5mu \tilde\chi}
\def\ueta{\lower 1.5ex\hbox{$\sim$}\mkern-13.5mu \eta}
\def\uetatilde{\lower 1.5ex\hbox{$\sim$}\mkern-13.5mu \tilde\eta}
\def\Psibar{\bar\Psi}
\def\uPsibar{\,\lower 1.2ex\hbox{$\sim$}\mkern-13.5mu \Psibar}
\def\upsi{\,\lower 1.5ex\hbox{$\sim$}\mkern-13.5mu \psi}
\def\uphi{\lower 1.5ex\hbox{$\sim$}\mkern-13.5mu \phi}
\def\uphione{\lower 1.5ex\hbox{$\sim$}\mkern-13.5mu \phi_1}
\def\uph2{\lower 1.5ex\hbox{$\sim$}\mkern-13.5mu \phi_2}
\def\psibar{\bar\psi}
\def\upsibar{\,\lower 1.5ex\hbox{$\sim$}\mkern-13.5mu \psibar}
\def\etabar{\bar\eta}
\def\uetabar{\,\lower 1.5ex\hbox{$\sim$}\mkern-13.5mu \etabar}
\def\chibar{\bar\chi}
\def\uchibar{\,\lower 1.5ex\hbox{$\sim$}\mkern-13.5mu \chibar}
\def\upsibarzero{\,\lower 1.5ex\hbox{$\sim$}\mkern-13.5mu \psibar^\zero}
\def\ulambda{\,\lower 1.2ex\hbox{$\sim$}\mkern-13.5mu \lambda}
\def\ulambdabar{\,\lower 1.2ex\hbox{$\sim$}\mkern-13.5mu \lambdabar}
\def\ulambdabarzero{\,\lower 1.2ex\hbox{$\sim$}\mkern-13.5mu \lambdabar^\zero}
\def\ulambdabarnew{\,\lower 1.2ex\hbox{$\sim$}\mkern-13.5mu \lambdabar^\new}
\def\D{{\cal D}}
\def\M{{\cal M}}
\def\N{{\cal N}}
\def\Dslash{\,\,{\raise.15ex\hbox{/}\mkern-12mu D}}
\def\Dbarslash{\,\,{\raise.15ex\hbox{/}\mkern-12mu {\bar D}}}
\def\delslash{\,\,{\raise.15ex\hbox{/}\mkern-9mu \partial}}
\def\delbarslash{\,\,{\raise.15ex\hbox{/}\mkern-9mu {\bar\partial}}}
\def\L{{\cal L}}
\def\hf{{\textstyle{1\over2}}}
\def\quarter{{\textstyle{1\over4}}}
\def\twe{{\textstyle{1\over12}}}
\def\eighth{{\textstyle{1\over8}}}
\def\fourth{\quarter}
\def\wb{Wess and Bagger}
\def\xibar{\bar\xi}
\def\ss{{\scriptscriptstyle\rm ss}}
\def\sc{{\scriptscriptstyle\rm sc}}
\def\uvcl{{\uv}^\cl}
\def\uAcl{\,\lower 1.2ex\hbox{$\sim$}\mkern-13.5mu A^{}_{\cl}}
\def\uAbarcl{\,\lower 1.2ex\hbox{$\sim$}\mkern-13.5mu A_{\cl}^\dagger}
\def\upsinew{{\upsi}^\new}
\def\ASDzero{{{\scriptscriptstyle\rm ASD}\zero}}
\def\SDzero{{{\scriptscriptstyle\rm SD}\zero}}
\def\SD{{\scriptscriptstyle\rm SD}}
\def\varthetabar{{\bar\vartheta}}
\def\three{{\scriptscriptstyle(3)}}
\def\dagthree{{\dagger\scriptscriptstyle(3)}}
\def\ld{{\scriptscriptstyle\rm LD}}
\def\vld{v^\ld}
\def\Dld{{\rm D}^\ld}
\def\Fld{F^\ld}
\def\Ald{A^\ld}
\def\Fbarld{F^{\dagger\scriptscriptstyle\rm LD}}
\def\Abarld{A^{\dagger\scriptscriptstyle \rm LD}}
\def\lambdald{\lambda^\ld}
\def\lambdabarld{\bar\lambda^\ld}
\def\psild{\psi^\ld}
\def\psibarld{\bar\psi^\ld}
\def\dsiginst{d\sigma_{\scriptscriptstyle\rm inst}}
\def\xione{\xi_1}
\def\xionebar{\bar\xi_1}
\def\xitwo{\xi_2}
\def\xitwobar{\bar\xi_2}
\def\thetatwo{\vartheta_2}
\def\thetatwobar{\bar\vartheta_2}
\def\Ltwo{\L_{\sst SU(2)}}
\def\Leff{\L_{\rm eff}}
\def\Laux{\L_{\rm aux}}
\def\oneloop{{\sst\rm 1\hbox{-}\sst\rm loop}}
\def\LSUtwo{{\cal L}_{\rm SU(2)}}
\def\Dhat{\hat\D}
\def\bkgd{{\sst\rm bkgd}}
\def\Lgft{{\cal L}_{\sst\rm g.f.t.}}
\def\Lghost{{\cal L}_{\sst\rm ghost}}
\def\Sinst{S_{\rm inst}}
\def\etal{{\rm et al.}}
\def\S{{\cal S}}
\def\L{ {\cal L}}
\def\C{ {\cal C}}
\def\N{ {\cal N}}
\def\calE{{\cal E}}
\def\lin{{\rm lin}}
\def\Tr{{\rm Tr}}
\def\mxth{\mathsurround=0pt }
\def\xversim#1#2{\lower2.pt\vbox{\baselineskip0pt \lineskip-.5pt
x  \ialign{$\mxth#1\hfil##\hfil$\crcr#2\crcr\sim\crcr}}}
\def\simgr{\mathrel{\mathpalette\xversim >}}
\def\simle{\mathrel{\mathpalette\xversim <}}
\def\slash{\llap /}
\def\lagr{{\cal L}}

\renewcommand{\a}{\alpha}
\renewcommand{\b}{\beta}
\renewcommand{\c}{\gamma}
\renewcommand{\d}{\delta}
\newcommand{\pa}{\partial}
\newcommand{\g}{\gamma}
\newcommand{\G}{\Gamma}
\newcommand{\e}{\epsilon}
\newcommand{\z}{\zeta}
\newcommand{\Z}{\Zeta}
\newcommand{\K}{\Kappa}
\renewcommand{\l}{\lambda}
\renewcommand{\L}{\Lambda}
\newcommand{\m}{\mu}
\newcommand{\n}{\nu}
\newcommand{\X}{\Chi}

\newcommand{\s}{\sigma}
\renewcommand{\S}{\Sigma}
\renewcommand{\t}{\tau}
\newcommand{\T}{\Tau}
\newcommand{\y}{\upsilon}
\newcommand{\Y}{\upsilon}
\renewcommand{\o}{\omega}
\newcommand{\q}{\theta}
\newcommand{\h}{\eta}
\newcommand{\cmap}{{$\bf c$} map}
\newcommand{\Ka}{K\"ahler} 
\renewcommand{\O}{{\Omega}}
\newcommand{\var}{\varepsilon}
%

\newcommand{\nd}[1]{/\hspace{-0.5em} #1}
\begin{titlepage}
\begin{flushright}
{\bf June 1998} \\ 
UW/PT 98-11 \\ 
hep-th/9806056 \\
\end{flushright}
\begin{centering}
\vspace{.2in}
{\large {\bf The BPS spectra of Two-Dimensional 
Supersymmetric \\ Gauge Theories with Twisted Mass Terms}}\\
\vspace{.4in}
 N. Dorey \\
\vspace{.4in}
Department of Physics, University of Washington, Box 351560   \\
Seattle, Washington 98195-1560, USA\\
\vspace{.2in}
and \\ 
\vspace{.2in}
Department of Physics, University of Wales Swansea \\
Singleton Park, Swansea, SA2 8PP, UK\\
\vspace{.4in}
{\bf Abstract} \\
\end{centering}
The vacuum structure and spectra of two-dimensional gauge theories with 
${\cal N}=(2,2)$ supersymmetry are investigated. These theories admit a 
twisted mass term for charged chiral matter multiplets. In the case of a $U(1)$ 
gauge theory with $N$ chiral multiplets of equal charge, an exact description of 
the BPS spectrum is obtained for all values of the twisted masses. The BPS 
spectrum has two dual descriptions which apply in the Higgs and Coulomb phases 
of the theory respectively.  The two descriptions are related by massive analog 
of mirror symmetry: the exact mass formula which is given by a one-loop calculation
in the Coulomb phase gives predictions for an infinite series of instanton 
corrections in the Higgs phase. The theory is shown to exhibit many phenomena 
which are usually associated with ${\cal N}=2$ theories in four dimensions. 
These include BPS-saturated dyons which carry both topological and Noether 
charges, non-trivial monodromies of the spectrum in the complex parameter space, 
curves of marginal stability on which BPS states can decay and strongly coupled 
vacua with massless solitons and dyons.


\end{titlepage}

\section{Introduction}
\paragraph{}

The purpose of this paper 
is to present some exact results for the mass spectrum of abelian 
gauge theories in two dimensions with ${\cal N}=(2,2)$ supersymmetry. 
In recent work \cite{hnh},  
Hanany and Hori introduced a new relevant parameter for these theories: 
a twisted mass for chiral superfields which corresponds to 
the expectation value of a background twisted chiral multiplet\footnote{Details 
of the multiplets of ${\cal N}=(2,2)$ supersymmetry in two dimensions 
and the corresponding superfields are reviewed in Section 2 below.}. 
The main new result presented below is an exact description of 
the spectrum of BPS states as a function of the twisted masses. 
Although most 
of the results presented here can easily be generalized to other 
${\cal N}=(2,2)$ theories, 
I will consider a model with gauge group $U(1)_{G}$ and $N$ chiral 
multiplets of equal charge. Without twisted masses this theory  
reduces to the supersymmetric $CP^{N-1}$ $\sigma$-model 
at low energy and its exact spectrum is well known \cite{KK,CV1,CV2}. 
In contrast, I will argue that the theory with non-zero twisted masses 
exhibits many phenomena which are new in two 
dimensions but familiar in the context of ${\cal N}=2$ theories 
in four dimensions \cite{SW1,SW2}. These include BPS dyons which carry 
both topological and Noether charges, a 
two-dimensional analog of the Witten effect \cite{WE}, 
non-trivial monodromies of the spectrum, 
curves of marginal stability on which BPS states can decay and  
strong-coupling vacua with massless solitons and dyons. The 
correspondence between two- and four-dimensional theories will be 
made precise below by identifying a complex curve whose periods 
govern the BPS spectra of both models.  
\paragraph{}
Two-dimensional ${\cal N}=(2,2)$ gauge theories 
have been studied extensively in the past because 
of their close relation to world-sheet conformal field theories which 
arise in compactifications of Type II string theory on Calabi-Yau 
manifolds and the phenomenon of mirror symmetry \cite{W1,MP}. Even with zero 
twisted masses, the model considered here is not conformally invariant 
but has a mass gap generated by strong quantum fluctuations in the 
infra-red. Nevertheless the model has a property which is a massive analog of 
mirror symmetry.  In two different regions of parameter space, the theory
is realized in a Higgs phase where the $U(1)$ 
gauge symmetry is spontaneously broken and a Coulomb phase with 
an unbroken gauge symmetry. An exact formula 
for BPS masses can be derived in the Coulomb phase by a one-loop calculation. 
The same formula applies in the Higgs phase 
where it predicts the exact numerical coefficients of an 
infinite series of instanton corrections. 
The main evidence in favour of the results presented here 
comes from a feature of the theory which has not been utilized before: 
the Higgs phase description of the theory 
is weakly coupled for large values of the twisted masses. 
These results are described in the remainder of this section 
while further details of the corresponding calculations 
appear in the subsequent sections. Theories with ${\cal N}=(2,2)$ supersymmetry 
have recently been discussed from a different point of view in \cite{L,P}         
\paragraph{}   
A gauge multiplet of ${\cal N}=(2,2)$ supersymmetry in two dimensions 
contains a complex scalar $\sigma$ which is the lowest component of 
a twisted chiral superfield $\Sigma$. The theory considered here also 
contains $N$ chiral superfields, $\Phi_{i}$  $i=1,2,\ldots,N$, each 
of charge $+1$ under $U(1)_{G}$, whose lowest components are complex 
scalars $\phi_{i}$. The parameters of the theory include a 
gauge coupling $e$, which has the dimensions of mass, as well as 
a Fayet-Iliopoulis (FI) parameter, $r$, and vacuum angle, $\theta$.  
The FI parameter and vacuum angle are dimensionless and it is convenient 
to combine them as a single complex parameter, 
$\tau=ir+\theta/2\pi$. As mentioned above, it is also possible to include 
a twisted mass $m_{i}$ for each chiral superfield $\Phi_{i}$. Only 
the differences between the twisted masses are physically significant:  
$\sum_{i=1}^{N}m_{i}$ can be set to zero by a linear shift in $\sigma$. 
In the absence of central charges, the ${\cal N}=(2,2)$ supersymmetry 
algebra has two abelian R-symmetries denoted $U(1)_{R}$ and $U(1)_{A}$. 
More generally, the supersymmetry algebra can be modified by including a central 
charge which breaks one of the two R-symmetries. In this case the spectrum of 
the corresponding theory can include massive BPS saturated states which lie in 
special, short representations of supersymmetry \cite{WO,W2}. 
In the following, the vacuum structure and the spectrum BPS states 
in the classical theory and in the corresponding quantum theory 
will be considered in turn.  
\subsubsection*{The classical theory}
\paragraph{}   
The classical theory with zero twisted masses has the maximal 
R-symmetry group, $U(1)_{R}\times U(1)_{A}$. The massless theory also has an 
$SU(N)$ global symmetry under which the chiral multiplets transform in the fundamental 
representation. For $r=0$, the theory has a classical Coulomb branch, with 
$\phi_{i}=0$ and $\sigma$ unconstrained. In these vacua the $U(1)_{G}$ gauge symmetry 
is unbroken and the gauge multiplet fields are classically massless. In contrast, 
for $r>0$, the D-term conditions for a supersymmetric vacuum set $\sigma=0$ and 
require that,   
\begin{equation}
\sum_{i=1}^{N}|\phi_{i}|^{2}=r
\label{dteq}
\end{equation}
The resulting space of gauge-inequivalent classical vacua is a copy of $CP^{N-1}$. 
At each point on this vacuum manifold at least one of the 
charged scalars, $\phi_{i}$, is non-zero and the $U(1)_{G}$ gauge-symmetry is 
spontaneously broken. This is the classical Higgs branch of the theory. 
The theory on the classical Higgs branch has $N-1$ 
massless chiral multiplets corresponding to the flat directions tangent to the 
vacuum manifold. The remaining degrees of freedom get masses of order $\sqrt{r}e$ 
by the Higgs mechanism. The effective theory for 
energies much less than this mass scale is a 
supersymmetric $\sigma$-model with target space $CP^{N-1}$. 
The coupling constant of the $\sigma$-model is related to the FI 
parameter of the underlying gauge theory as $g \sim 1/\sqrt{r}$. The $CP^{N-1}$ 
target space is covered by $N$ different coordinate patches, 
${\cal P}_{j}$ with $j=0,1,\ldots,N$, each with 
$N-1$ complex coordinates $w^{(j)}_{i}=\phi_{i}/\phi_{j}$ with $i\neq j$.  
The patch ${\cal P}_{j}$ covers the complement in $CP^{N-1}$ of the submanifold defined by 
$\phi_{j}=0$. 
\paragraph{}
The inclusion of twisted masses,  
with $m_{i}\neq m_{j}$ for each $i$ and $j$, changes the classical analysis given above in 
several ways. First, as shown in Section 3 below, 
the continuous vacuum degeneracy discovered above is lifted and the model has 
$N$ isolated supersymmetric vacua. Specifically there is exactly one vacuum ${\cal V}_{i}$ 
in each of the coordinate patches ${\cal P}_{i}$ defined above. In each vacuum ${\cal V}_{i}$ 
the low-energy effective theory is a variant of the supersymmetric 
$CP^{N-1}$ $\sigma$-model with explicit mass terms for each of the $N-1$ gauge-invariant 
fields $w^{(i)}_{j}=\phi_{j}/\phi_{i}$ with $j\neq i$ (and their superpartners). Second, 
the $SU(N)$ global symmetry of the massless theory is explicitly 
broken to its maximal abelian subgroup 
$U(1)^{N-1}=(\otimes_{i=1}^{N} U(1)_{i})/U(1)_{G}$. Here $U(1)_{i}$ denotes the global symmetry 
with generator $S_{i}$ under which $\phi_{j}$ has charge $+1$ if $j=i$ and zero otherwise. 
Finally, in the theory with non-zero twisted masses, 
the R-symmetry $U(1)_{A}$ is broken down to a discrete subgroup, $Z_{2}$. As mentioned above 
this permits a non-zero central charge to appear in the supersymmetry algebra, which is 
a necessary condition for the existence of BPS states. This possibility 
is analysed in detail in Section 4. It is shown that the classical 
spectrum includes three different kinds of BPS states,   
\paragraph{}
{\bf 1:} In the vacuum ${\cal V}_{i}$, the 
BPS spectrum includes the 
elementary quanta of the $N-1$ $\sigma$-model fields $w^{(i)}_{j}$ with 
$j\neq i$ (and superpartners). These states carry the global $U(1)$ Noether charges 
$\vec{S}=(S_{1},S_{2},\ldots S_{N})$. Including the 
states from each vacuum the spectrum includes BPS states 
with all $N(N-1)$ possible charge vectors of the form $\pm(0,\ldots,+1,
\ldots,-1,\ldots,0)$. 
\paragraph{}
{\bf 2:} A two-dimensional theory with isolated vacua can have 
topologically stable solitons. 
These are classical field configurations which asymptote to one vacuum, ${\cal V}_{L}$, 
at left spatial infinity and a different vacuum, ${\cal V}_{R}$, at right spatial infinity. 
In this connection it is useful to define topological 
charges $T_{i}$ which are equal to $+1$ if ${\cal V}_{R}={\cal V}_{i}$, $-1$ if  
is ${\cal V}_{L}={\cal V}_{i}$, and zero otherwise. 
The theory with twisted masses has Bogomol'nyi saturated solitons which 
interpolate between each pair of vacua,  
${\cal V}_{i}$ and ${\cal V}_{j}$ with $i\neq j$. The solitons yield $N(N-1)$ 
BPS multiplets of ${\cal N}=(2,2)$ supersymmetry.   
These states carry the topological charges $\vec{T}=(T_{1},T_{2},\ldots T_{N})$. 
As for the Noether charges, the spectrum includes all states with charge vectors of the 
form $\pm(0,\ldots,+1,\ldots,-1,\ldots,0)$. 
\paragraph{}
{\bf 3:} The solitons of the model have a feature which is unusual for 
topologically stable kinks in two dimensions: they have an internal degree 
of freedom corresponding to global $U(1)$ rotations. Specifically, the 
soliton which interpolates between the vacua ${\cal V}_{i}$ and 
${\cal V}_{j}$ transforms under the global $U(1)$ symmetry with generator 
$S_{j}-S_{i}$. Quantizing this degree 
of freedom yields an infinite tower of `dyons' which carry both 
Noether and topological charges. 
For each allowed topological charge vector $\vec{T}$, the spectrum contains 
BPS dyons with Noether charge vectors $\vec{S}=S\vec{T}$ where $S$ can be any 
integer. These dyons exhibit an exact analog of the Witten effect \cite{WE} 
in four-dimensions; in the presence of a non-zero vacuum angle 
their global $U(1)$ charge is shifted by an amount $\theta/2\pi$. Similar 
two-dimensional dyons have been studied previously in \cite{QK}.      
\paragraph{}
The masses of all the 
states described above are given by a BPS mass formula $M=|Z|$ with, 
\begin{equation}
Z=-i\vec{m}\cdot(\vec{S}+\tau\vec{T}) 
\label{ccharge1}
\end{equation}
where $\vec{m}=(m_{1},m_{2},\ldots,m_{N})$. 
The classical BPS spectrum described above coincides 
exactly with the classical spectrum of massive BPS states 
of ${\cal N}=2$ supersymmetric $SU(N)$ Yang-Mills theory 
in four dimensions. In particular, in the spectrum of the two-dimensional 
theory there is a one-to-one correspondence between elementary particles and solitons 
of exactly the same form as that noted by Goddard, Nuyts and Olive \cite{GNO} for 
the classical $D=4$ gauge theory. The global $SU(N)$ symmetry of the two-dimensional theory 
corresponds to the gauge symmetry of the four-dimensional theory. 
The two-dimensional twisted masses $m_{i}$ correspond to the eigenvalues of the 
adjoint scalar VEV which breaks $SU(N)$ down to $U(1)^{N-1}$ on the Coulomb 
branch of the four-dimensional theory. Hence the components of 
$\vec{S}$ correspond to the abelian electric charges in $D=4$ 
and those of $\vec{T}$ to the magnetic charges. The complex coupling 
$\tau=ir+\theta/2\pi$ is mapped onto the complexified gauge coupling
$\tau_{4D}=4\pi i/g_{4D}^{2}+\theta_{4D}/2\pi$ of the four-dimensional 
theory. The correspondence between the couplings and symmetries of the 
two theories was noted in \cite{hnh} and explained in the context of 
intersecting D-branes in type IIA string theory. It would be interesting 
to try an understand the relation described here 
between the spectra of the two theories in the same way.   
  
\subsubsection*{The quantum theory}
\paragraph{}
The classical theory described above is modified in several ways 
by quantum effects. The FI parameter runs logarithmically at one-loop 
and is traded for an RG-invariant scale $\Lambda$ by dimensional 
transmutation. In addition, the $U(1)_{A}$ R-symmetry of the 
theory without twisted masses is broken by an anomaly to 
a residual discrete symmetry $Z_{2N}$. This means that 
the bare $\theta$-parameter can be set to zero by a 
chiral rotation of the fields (a phase rotation of the twisted masses 
is also necessary if they are non-zero). The physical parameters of the 
quantum theory are the gauge coupling $e$, the dynamical scale 
$\Lambda$ and the twisted masses $m_{i}$. The are two 
different regions of this parameter space 
in which the theory has a weakly-coupled description and its 
properties can be analysed reliably. 
\paragraph{}
${\bf a: e>>|m_{i}-m_{j}|>>\Lambda}$ In this regime the low-energy theory 
is described by the classically massive version of the supersymmetric 
$CP^{N-1}$ $\sigma$-model discussed in the previous section. 
Although the $\sigma$-model is asymptotically free its coupling  
only runs for energy scales larger than the masses $|m_{i}-m_{j}|$. 
Hence the low-energy theory is weakly-coupled as long as 
$|m_{i}-m_{j}|>>\Lambda$. In this region of parameter space the 
$U(1)_{G}$ gauge symmetry is spontaneously broken and the BPS spectrum 
is qualitatively similar to that of the classical theory described above. 
The classical spectrum is corrected by one-loop effects which are calculated 
explicitly for the $N=2$ case in Section 5. There are also non-perturbative 
corrections from all numbers of $\sigma$-model instantons.    
\paragraph{}
${\bf b: e<<\Lambda}$ The theory in this regime consists a light  
gauge multiplet weakly coupled to massive chiral multiplets \cite{W1}. 
In particular the dimensionful gauge coupling is much 
smaller then the other relevant mass scales and 
the model can be analysed using ordinary 
perturbation theory. Note that this is a completely different expansion  
to the perturbation in the $\sigma$-model coupling considered above above. 
A one-loop calculation suffices to show that the theory has $N$ 
isolated vacua, each with unbroken $U(1)_{G}$ gauge symmetry \cite{W1}. 
The BPS spectrum consists entirely of 
solitons which are charged under $U(1)_{G}$ and 
interpolate between different vacua. In the absence of 
twisted masses the solitons lie in multiplets of the unbroken 
$SU(N)$ global symmetry. Introducing small twisted masses breaks this symmetry and 
introduces mass splittings between degenerate states.     
\paragraph{}
Superficially, it appears that the descriptions of the theory in these 
two regions of parameter space are completely different. 
For $e>>\Lambda$, the gauge symmetry is spontaneously 
broken and the theory is in a Higgs phase. In contrast, for $e<<\Lambda$, 
$U(1)_{G}$ is unbroken and, adopting the terminology of four 
dimensional gauge theories, the theory is in its Coulomb phase\footnote{In two dimensions 
a Coulomb interaction between two charges leads to a confining linear potential. However 
in the supersymmetric theory considered here the gauge multiplet gets a mass from quantum 
effects and the Coulomb interaction is screened \cite{W2,ALD1}. Thus there are no long-range 
gauge interactions in either phase.}. 
Despite these differences, some features of the theory remain the same in 
both phases. The simplest example is the Witten index which counts 
the number of supersymmetric vacua weighted by fermion number. 
Two-dimensional theories with ${\cal N}=(2,2)$ supersymmetry also 
have another supersymmetric index which was introduced by Cecotti, 
Fendley, Intriligator and Vafa (CFIV) in \cite{CFIV}. This index is invariant under D-term 
variations of the superspace Lagrangian. 
This is a refinement 
of the Witten index as the latter is invariant under both F- and D-term 
variations (subject to certain boundary conditions). While only the vacuum states 
of the theory contribute to the Witten 
index, the CFIV index is sensitive to all states in short representations of 
supersymmetry. In fact, part of the information contained in 
the index is the mass and degeneracy of each BPS saturated state \cite{CV2}. 
\paragraph{}
The CFIV index is relevant in the present context because 
the gauge kinetic term can be written as a D-term in ${\cal N}=(2,2)$ superspace \cite{W1}.  
In addition, the fields can be rescaled so that the gauge coupling only appears in 
this term in the action. It follows from the above discussion 
that the masses of BPS states are independent of $e$. For this reason we can calculate 
BPS masses using the weakly-coupled Coulomb phase description of the theory which is valid 
for $e<<\Lambda$ and apply the results for all values of $e$. This calculation 
is described in Section 6. The result is that, for all values of the parameters, 
the mass of a BPS state with global Noether charge $\vec{S}$ and topological charge 
$\vec{T}$ is given by $M=|Z|$ with 
\begin{equation}    
Z=-i(\vec{m}\cdot\vec{S}+\vec{m}_{D}\cdot\vec{T})    
\label{exact1}
\end{equation}
where $\vec{m}_{D}=(m_{D1},m_{D2},\ldots,m_{DN})$ and
$m_{Di}=Ne_{i}-\sum_{j=1}^{N}m_{j}\log(e_{i}+m_{j})$. Here $e_{i}$, with $i=1,2,\ldots,N$ 
denote the roots of the polynomial equation,
\begin{equation}
\prod_{i=1}^{N}(x+m_{i})-\tilde{\Lambda}^{N}=0
\label{exact2}
\end{equation}
where $\tilde{\Lambda}=\Lambda\exp(-1+i\theta/N)/2$. 
For $|m_{i}-m_{j}|>>\tilde{\Lambda}$, (\ref{exact1}) can be compared directly with  
semiclassical results obtained using the massive $\sigma$-model description 
of the theory. The exact soliton and dyon masses have a non-trivial expansion in the 
small parameters $\Lambda/|m_{i}-m_{j}|$ which corresponds to the weak-coupling expansion 
of the $\sigma$-model. In general the expansion contains terms corresponding to one-loop 
perturbation theory as well as an infinite series of corrections which can be interpreted as 
coming from $\sigma$-model instantons. In the simplest case $N=2$, the exact 
formula predicts a one-loop correction to the classical spectrum (\ref{ccharge1}) which 
is equivalent to the replacement $\tau\rightarrow \tau_{\rm eff}+1$ where 
$\tau_{\rm eff}=i\log(m/\tilde{\Lambda})/\pi$ and $m=m_{1}-m_{2}$. 
In Section 5 this result is tested against an explicit semiclassical calculation 
of quantum corrections to the soliton mass. The above results also predict 
two-dimensional analogs for several phenomena which occur in four-dimensional 
gauge theories with ${\cal N}=2$ supersymmetry:     
\paragraph{}
{\bf 1:} The branch-cuts appear in the exact formula for $\vec{m}_{D}$ 
implies that the BPS spectrum undergoes monodromies in the complex parameter space. 
In the $N=2$ case, with $m=m_{1}-m_{2}$, there is a 
non-trivial monodromy around the point at infinity in the 
complex $m$-plane. In this case, the charge vectors $\vec{S}$ and $\vec{T}$ transform as  
\begin{eqnarray}
\vec{S}\rightarrow \vec{S}-2\vec{T}  & \qquad{} \qquad{} \qquad{} & \vec{T}\rightarrow \vec{T}
\label{monod1}
\end{eqnarray}  
In Sections 5 this effect is derived explicitly using weak-coupling methods.  
\paragraph{}
{\bf 2:} There are submanifolds of the parameter space on which the roots 
of equation (\ref{exact2}) become degenerate. On these submanifolds BPS 
states become massless. In particular, solitons and dyons which are very 
massive at weak coupling can become massless at strong coupling for some 
values of the parameters.    
\paragraph{}
{\bf 3:} There can be curves of marginal stability (CMS) on which BPS states can 
decay. Typically these curves will have real codimension one 
in the parameter space and therefore can disconnect different regions of 
this space. These curves play an important 
role in resolving the remaining discrepancies 
between the BPS spectra in different regions of parameter space. For $|m_{i}-m_{j}|>>\Lambda$ 
the semiclassical spectrum described above includes 
an infinite number of stable BPS states. In contrast, for 
$|m_{i}-m_{j}|<<\Lambda$, the BPS spectrum 
should be close to that of the supersymmetric $CP^{N-1}$ $\sigma$-model 
which has only a finite number of such states. This disparity can be resolved 
if the regions of large and small twisted mass are separated 
by a curve of marginal stability. 
Some explicit checks that the required CMS is present in the $N=2$ case 
are performed in Section 6. 
\paragraph{}   
These effects suggest that there is a correspondence between the 
two-dimensional theory with twisted masses and an
${\cal N}=2$ supersymmetric gauge theory in four-dimensions 
which holds at the quantum level. 
This can be made precise by noting that the soliton masses implied by 
the exact formula (\ref{exact1}) correspond to the periods of the following 
degenerate elliptic curve;
\begin{equation}
(t-\tilde{\Lambda}^{N})\left(t-\prod_{i=1}^{N}(x+m_{i})\right)=0
\label{curvec1}
\end{equation}
This is the same curve which describes an ${\cal N}=2$ gauge theory in 
four-dimensions with gauge-group $SU(N)$ and $N$ hypermultiplets in the 
fundamental representation. Specifically, the four-dimensional theory is 
at a particular point on its Coulomb branch which is 
the root of the baryonic Higgs branch \cite{APS}. One application 
of this correspondence is that, at least for the $N=2$ case, 
the existence of the CMS described in {\bf 3} above 
can be deduced from the known behaviour of the four-dimensional theory \cite{BF2,BF}.  
The condition (\ref{exact2}) for a supersymmetric vacuum in the two-dimensional Coulomb 
phase can be interpreted as the conditions obeyed by the singular points 
on the complex manifold (\ref{curvec1}). 
This phenomenon seems to be a massive generalization of the description 
of mirror symmetry between ${\cal N}=(2,2)$ conformal theories given in \cite{MP}.  
Finally it would be very interesting to find an explanation for the results 
described above in the context of string theory. As discussed in \cite{hnh}, 
configurations of intersecting D2, D4 and NS5 branes in type IIA string theory, 
which become M2 and M5 branes in M-theory, provide a natural way to 
relate world-volume gauge theories in two- and four-dimensions.     

\section{Fields and Symmetries and Dimensional Reduction}
\paragraph{}
This section contains a review of the basic features of theories with 
${\cal N}=(2,2)$ supersymmetry in two dimensions.  
These theories were studied in detail by Witten in \cite{W1} and the 
presentation given here closely follows this reference. With a few 
exceptions, the notation and conventions adopted below are those of 
\cite{W1}. This section also includes a review of BPS saturated solitons 
in two dimensions.   
\paragraph{}
Theories with ${\cal N}=(2,2)$ supersymmetry in two dimensions (2D) may be obtained 
by dimensional reduction of four-dimensional theories with ${\cal N}=1$ supersymmetry. 
Specifically, we will start in four-dimensional Minkowski space with coordinates $X_{m}$ 
$m=0,1,2,3$ and obtain a two-dimensional theory by considering field configurations 
which are independent of $X_{1}$ and $X_{2}$. The two-dimensional spacetime coordinate is 
denoted $x_{\mu}$ with $\mu=0,1$ where $x_{0}=X_{0}$ and $x_{1}=X_{3}$.   
A four-vector $A_{m}$ reduces to a two-vector $a_{\mu}$ and two real scalars.  
A left-handed Weyl spinor $\psi_{\alpha}$ in four-dimensions yields a complex spinor in 
two dimensions with components $(\psi_{-},\psi_{+})=(\psi_{1},\psi_{2})$. Similarly, 
a right-handed Weyl spinor $\bar{\psi}_{\dot{\alpha}}$ 
in four-dimensions yields a complex spinor in 
two dimensions with components $(\bar{\psi}_{-},\bar{\psi}_{+})=(\bar{\psi}_{\dot{1}},
\bar{\psi}_{\dot{2}})$. The components of $\psi_{\alpha}$ and $\bar{\psi}_{\dot{\alpha}}$ 
can be combined to make a two-dimensional Dirac spinor $\Psi$ and its charge conjugate 
$\bar{\Psi}$. However, following the notation of \cite{W1} we will mostly work in terms of the 
components $\psi_{\pm}$ and $\bar{\psi}_{\pm}$ and use the four-dimensional 
notation for spinors with summation over $+$ and $-$ components for repeated 
indices\footnote{Like the 4D spinor index, the 
2D Dirac index is raised and lowered with the 
antisymmetric tensor $\epsilon^{12}=-\epsilon_{12}=1$. Thus 
$(\psi^{-},\psi^{+})=(\psi^{1},\psi^{2})$  with $\psi^{-}=\psi_{+}$, $\psi^{+}=-\psi_{-}$ 
and similar relations for $\bar{\psi}^{\pm}$.}.   
For example we have 
$\psi^{\alpha}\psi_{\alpha}=\psi^{-}\psi_{-}+\psi^{+}\psi_{+}=2\psi_{+}\psi_{-}$.    
\paragraph{}
The ${\cal N}=1$ superalgebra in four dimensions contains left- and right-handed Weyl supercharges 
$Q_{\alpha}$ and $\bar{Q}_{\dot{\alpha}}$ with anti-commutator,     
\begin{equation}
\{Q_{\alpha},\bar{Q}_{\dot{\alpha}}\}=2\sigma^{m}_{\alpha\dot{\alpha}}P_{m}    
\label{4Dsusyalg}
\end{equation}
On dimensional reduction to two dimensions, we obtain the ${\cal N}=(2,2)$ supersymmetry algebra 
for two Dirac supercharges $Q_{\pm}$ and $\bar{Q}_{\pm}$,   
\begin{equation}
\begin{array}{ll}
\{Q_{+},\bar{Q}_{+}\}=-2(p_{0}+p_{1})\qquad  &  \qquad{} 
\{Q_{-},\bar{Q}_{-}\}=-2(p_{0}-p_{1})  \\ 
\{Q_{-},\bar{Q}_{+}\}= \,\,\,\,2Z  \qquad{} & \qquad{} 
\{Q_{+},\bar{Q}_{-}\}=\,\,\, \,2\bar{Z}  \\
\end{array}
\label{susyalg} 
\end{equation}
where $p_{0}=P_{0}$, $p_{1}=P_{3}$. $p_{0}\pm p_{1}$ correspond to 
right and left moving momentum in two dimensions.  The components of four-momentum 
in the reduced directions yield a complex central charge $Z=P_{1}-iP_{2}$ which 
will play an important role in the following. 
\paragraph{}
The two-dimensional supersymmetry algebra (\ref{susyalg}) inherits a $U(1)$ 
R-symmetry from the ${\cal N}=1$ superalgebra in four dimensions under 
which $Q_{\pm}$  has charge $+1$ and $\bar{Q}_{\pm}$
has charge $-1$. We will call this symmetry $U(1)_{R}$. If the central charge vanishes, 
then the two-dimensional superalgebra has an additional R-symmetry, 
denoted $U(1)_{A}$, which corresponds to spatial 
rotations in the two reduced dimensions. The $U(1)_{R}\times U(1)_{A}$ charges of the 
supersymmetry generators can be represented as    
\begin{equation}
\begin{array}{cc}
\bar{Q}_{+}\,\,&\,\,Q_{-}\\[0.4cm]
\bar{Q}_{-}\,\,&\,\,Q_{+} 
\end{array}
\end{equation}
\paragraph{}
where generators in the (left-) right-handed column have $U(1)_{R}$ charge ($-1$) $+1$. 
Generators in the (bottom) top row have $U(1)_{A}$ charge ($-1$) $+1$. 
\paragraph{}
In addition to the continuous 
$R$-symmetries described above the ${\cal N}=(2,2)$ supersymmetry algebra 
also has a discrete mirror automorphism which interchanges the supercharges $Q_{+}$ and 
$\bar{Q}_{+}$ and does not act on $Q_{-}$ and $\bar{Q}_{-}$. This symmetry also 
interchanges the R-symmetry groups $U(1)_{R}$ and $U(1)_{A}$ Clearly, 
this is only an automorphism of the algebra (\ref{susyalg}), 
if the central charge $Z$ vanishes. 
If $Z\neq 0$ the same transformation maps (\ref{susyalg}) to a 
mirror algebra in which a different central charge 
appears as the 
anti-commutator of $Q_{-}$ and $Q_{+}$.  
The new central charge breaks $U(1)_{R}$ but leaves $U(1)_{A}$ is unbroken.  
\paragraph{}     
Multiplets of ${\cal N}=1$ supersymmetry in four dimensions 
yield multiplets of ${\cal N}=(2,2)$ 
supersymmetry after dimensional reduction to two dimensions. However, 
as we discuss below, not all ${\cal N}=(2,2)$ multiplets can be 
obtained in this way. 
Following Witten \cite{W1}, it is convenient to start from an  
${\cal N}=1$ superspace in four dimensions with coordinates $X_{m}$, 
$\theta_{\alpha}$, $\bar{\theta}_{\dot{\alpha}}$. 
The four-dimensional chiral multiplet correponds an 
${\cal N}=1$ chiral superfield $\Phi(X,\theta,\bar{\theta})$ which obeys the constraint 
$\bar{D}_{\dot{\alpha}}\Phi=0$. In the standard notation of Wess and Bagger \cite{WB},  
this superfield has the component expansion,  
\begin{equation}
\Phi=\phi(Y)+\sqrt{2}\theta^{\alpha}\psi_{\alpha}(Y)+\theta^{\alpha}\theta_{\alpha}F(Y)
\label{chiralsf}
\end{equation} 
where $Y^{m}=X^{m}+i\theta^{\alpha}\sigma_{\alpha\dot{\alpha}}^{m}\bar{\theta}^
{\dot{\alpha}}$. The component fields include a complex scalar $\phi$, a 
left-handed Weyl fermion $\psi_{\alpha}$ and a complex auxiliary field $F$. 
Similarly, the anti-chiral superfield $\bar{\Phi}$ obeys the constraint 
$D_{\alpha}\bar{\Phi}=0$ and its component fields are the charge 
conjugate degrees of freedom $\bar{\phi}$, $\bar{\psi}_{\dot{\alpha}}$ and $\bar{F}$. 
\paragraph{}  
As above, we dimensionally reduce by considering only superfield configurations 
which are independent of $X_{1}$ and $X_{2}$. In the notation introduced above, this 
yields a two-dimensional chiral superfield $\Phi(x,\theta,\bar{\theta})$ 
which obeys the constraints $\bar{D}_{-}\Phi= \bar{D}_{+}\Phi=0$. The two-dimensional 
anti-chiral superfield obeys the constraints  $D_{-}\bar{\Phi}= D_{+}\bar{\Phi}=0$. 
Thus the four-dimensional (anti-)chiral multiplet reduces to a two-dimensional 
(anti-)chiral multiplet with the same scalar field content and a two-dimensional Dirac 
fermion with components $\psi_{\pm}$ ($\bar{\psi}_{\pm}$). The $U(1)_{R}\times U(1)_{A}$ 
charges of the dynamical fields can be represented as, 
\begin{equation}
\begin{array}{cccc}
& \bar{\psi}_{+}\,\, &\,\,\psi_{-} & \\[0.2cm]
\bar{\phi}\,\, & &  & \,\,  \phi \\[0.2cm]
& \bar{\psi}_{-}\,\,&\,\,\psi_{+} &
\end{array}
\end{equation}
where, as before, columns (rows) correspond to $U(1)_{R}$ ($U(1)_{A}$) charge. 
In this case the $U(1)_{R}$ ($U(1)_{A}$) charges ranges from $-2$ to $+2$ 
($-1$ to $+1$). In fact the assignment of $U(1)_{R}$ charges is more subtle 
than indicated above: it is possible to redefine the R-symmetry generators by 
adding to them the generators of other global symmetries \cite{W1}. However this 
will not play an important role in the following.      
\paragraph{}
The two basic invariant Lagrangians for two-dimensional 
chiral superfields $\Phi_{i}$ are obtained by 
dimensional reduction of their counterparts in four dimensions. These are the $D$-term 
Lagrangian\footnote{The conventions for superspace integration are those of \cite{WB}. 
In 2D notation these read $d^{2}\theta=d\theta_{-}d\theta_{+}/2$, 
$d^{2}\bar{\theta}=d\bar{\theta}_{-}d\bar{\theta}_{+}/2$ and $d^{4}\theta=d^{2}\theta 
d^{2}\bar{\theta}$}
\begin{equation}
{\cal L}_{D}=\int\,d^{4}\theta\, K(\Phi_{i},\bar{\Phi}_{i})
\label{Dterm}
\end{equation} 
where the \Ka\  potential, $K$, is a real function of 
$\Phi_{i}$ and $\bar{\Phi}_{i}$ and the F-term Lagrangian,
\begin{equation}
{\cal L}_{F}=\int\,d^{2}\theta\, W(\Phi_{i}) \, \, + \int\,
d^{2}\bar{\theta}\, \bar{W}(\bar{\Phi}_{i})
\label{Fterm}
\end{equation}
where the superpotential $W$ is holomorphic in $\Phi_{i}$. Both the D-term and the 
F-term are invariant under $U(1)_{A}$. The F-term is only invariant under $U(1)_{R}$ 
if the superpotential has charge $+2$ under this symmetry (modulo the subtlety about 
possible redefinitions of the $U(1)_{R}$ generator mentioned above).   
The bosonic terms in 
${\cal L}_{B}+{\cal L}_{F}$ are given by, 
\begin{equation}
{\cal L}_{\rm Bose}=g_{i\bar{j}}\left(-\partial_{\mu}\phi^{i}
\partial^{\mu}\bar{\phi}^{j}+F^{i}\bar{F}^{j}\right)+\left(
F^{i}\frac{\partial{ W}}{\partial\phi^{i}} + \bar{F}^{i}
\frac{\partial\bar{ W}}{\partial\bar{\phi}^{i}}\right)    
\label{lbosea}
\end{equation}
where the \Ka\ metric $g_{i\bar{j}}$ is defined as, 
\begin{equation} 
g_{i\bar{j}}=\frac{\partial^{2}  K}{\partial \phi^{i}\partial\bar{\phi}^{j}} 
\label{kmetric0}
\end{equation} 
\paragraph{}
The ${\cal N}=1$ gauge multiplet in four dimensions consists of the gauge field 
$V_{m}$, right- and left-handed Weyl spinors\footnote{The fermion field $\chi_{\alpha}$ 
introduced above is related to the fermion field  $\lambda_{\alpha}$ of Wess and Bagger by 
$\chi_{\alpha}=-\sqrt{2}i\lambda_{\alpha}$. The unconventional normalization is chosen so 
that the formalism is symmetric between chiral 
and twisted chiral supermultiplets which will be defined below}
$\chi_{\alpha}$ and $\bar{\chi}_{\dot{\alpha}}$ and a real auxiliary field $D$. 
These fields are components 
of a real superfield $V(X,\theta,\bar{\theta})$. After imposing the Wess-Zumino gauge 
condition, $V$ has the expansion,
\begin{equation}
V=-\theta^{\alpha}\sigma^{m}_{\alpha\dot{\alpha}}\bar{\theta}^{\dot{\alpha}}V_{m}-
\frac{1}{\sqrt{2}}\left(\theta^{\alpha}\theta_{\alpha}\bar{\theta}_{\dot{\alpha}}
\bar{\chi}^{\dot{\alpha}}
-\bar{\theta}_{\dot{\alpha}}\bar{\theta}^{\dot{\alpha}}\theta^{\alpha}\chi_{\alpha}\right)
+\frac{1}{2}\theta^{\alpha}\theta_{\alpha}\bar{\theta}_{\dot{\alpha}}
\bar{\theta}^{\dot{\alpha}}D
\label{wzgauge}
\end{equation}
After dimensional reduction, the four dimensional gauge field $V_{m}$ yields a 
two-dimensional gauge field $v_{\mu}$ with $v_{0}=V_{0}$ and $v_{1}=V_{3}$  
and a complex scalar $\sigma=V_{1}-iV_{2}$. The Weyl fermions 
in the four-dimensional gauge multiplet reduce to 
two Dirac fermions with components $\chi_{\pm}$ 
and $\bar{\chi}_{\pm}$ respectively. 
The $U(1)_{R}\times U(1)_{A}$ 
charges of these fields are, 
\begin{equation}
\begin{array}{ccc}
& \sigma & \\[0.2cm]
\bar{\chi}_{+}\,\,& & \,\,\chi_{-}  \\
   & v_{\mu} & \\
 \bar{\chi}_{-}\,\,& & \,\,\chi_{+} \\
&\bar{\sigma} &
\end{array}
\end{equation}
A chiral multiplet can be coupled to the gauge field with the minimal coupling 
prescription, 
\begin{equation}
\Phi \rightarrow \exp(V)\Phi
\label{mincoupling}
\end{equation} 
The resulting Lagrangian for the component fields is given in in Appendix A. 
\paragraph{}
In two dimensions it also possible to construct ${\cal N}=(2,2)$ 
multiplets, known as twisted chiral multiplets \cite{G}, 
which are chiral with repect to supercharges $Q_{-}$ and $\bar{Q}_{+}$. 
The corresponding twisted chiral superfield 
$\Lambda(x,\theta,\bar{\theta})$ obeys the constraints 
$D_{-}\Lambda=\bar{D}_{+}\Lambda=0$. 
This possiblility, which has no analog in four-dimensions, is closely related to the 
mirror automorphism of 
the supersymmetry algebra (\ref{susyalg}) which interchanges the supercharges $Q_{+}$ 
and $\bar{Q}_{+}$. To exhibit the symmetry between chiral and twisted chiral superfields 
it is convenient to introduce a twisted chiral superspace notation in which the fermionic 
coordinates $\theta_{+}$ and $\bar{\theta}_{+}$ are interchanged. Thus we define twisted 
fermionic coordinates ${\vartheta}_{\alpha}$ and  $\bar{{\vartheta}}_{\dot{\alpha}}$ 
with $({\vartheta}_{1},{\vartheta}_{2})=(\theta_{-},\bar{\theta}_{+})$ and 
$(\bar{{\vartheta}}_{\dot{1}}, \bar{{\vartheta}}_{\dot{2}})=
(\bar{\theta}_{-},\theta_{+})$. The twisted chiral superfield $\Lambda(x,\theta, 
\bar{\theta})$ has the expansion, 
\begin{equation}
\Lambda=\lambda({\cal Y})+\sqrt{2}{\vartheta}^{\alpha}\tilde{\rho}_{\alpha}({\cal Y})+
{\vartheta}^{\alpha}{\vartheta}_{\alpha}E({\cal Y})
\label{tchiralsf}
\end{equation} 
where ${\cal Y}^{m}=X^{m}+i{\vartheta}^{\alpha}
\sigma_{\alpha\dot{\alpha}}^{m}\bar{{\vartheta}}^{\dot{\alpha}}$. 
The component fields include a complex scalar $\lambda$, a 
2D fermion $\tilde{\rho}_{\alpha}$ with 
$(\tilde{\rho}_{1},\tilde{\rho}_{2})=(\rho_{-},\bar{\rho}_{+})$  
and a complex auxiliary field $E$. Similarly 
the twisted anti-chiral superfield $\bar{\Lambda}$ obeys the constraint 
$\bar{D}_{-}\bar{\Lambda}=D_{+}\bar{\Lambda}=0$ and has component fields $\bar{l}$, 
$\bar{\tilde{\rho}}_{\dot{\alpha}}$ and $\bar{L}$ where $(\bar{\tilde{\rho}}_{\dot{1}},
\bar{\tilde{\rho}}_{\dot{2}})=(\bar{\rho}_{-},\rho_{+})$. 
The $U(1)_{R}\times U(1)_{A}$ 
charges of the dynamical fields of these multiplets are given as, 
\begin{equation}
\begin{array}{ccc}
& \lambda & \\[0.2cm]
\bar{\rho}_{+}\,\,& & \,\,\rho_{-}  \\[0.2cm]
 \bar{\rho}_{-}\,\,& & \,\,\rho_{+} \\[0.2cm]
&\bar{\lambda} &
\end{array}
\end{equation}
An important example of a twisted chiral multiplet is the multiplet associated with the 
gauge field strength. 
The abelian field strength $f=\epsilon^{\mu\nu}\partial_{\mu}v_{\nu}$ is contained in a 
gauge-invariant superfield $\Sigma=\bar{D}_{+}D_{-}V$ whose lowest 
component is the complex scalar $\sigma$. 
By construction $\Sigma$ obeys 
$\bar{D}_{+}\Sigma=D_{-}\Sigma=0$ and is therfore twisted chiral. 
In addition to $\sigma$, the twisted chiral 
multiplet also contains the fermion components $\chi_{-}$ and 
$\bar{\chi}_{+}$ and the complex field $S=D-if$. With the chosen normalization 
for the component fields (see footnote $5$ above), $\Sigma$ 
has an expansion of the 
standard form (\ref{tchiralsf}), 
\begin{equation}
\Sigma=\sigma({\cal Y})+\sqrt{2}{\vartheta}^{\alpha}\tilde{\chi}_{\alpha}({\cal Y})+
{\vartheta}^{\alpha}{\vartheta}_{\alpha}S({\cal Y})
\label{tchiralsf2}
\end{equation} 
where, in the twisted superspace notation introduced above, the Dirac fermion 
$\tilde{\chi}_{\alpha}$ has components 
$(\tilde{\rho}_{1},\tilde{\rho}_{2})=(\rho_{-},\bar{\rho}_{+})$  
Similarly one can define a twisted 
anti-chiral multiplet which contains the complex scalars 
$\bar{\sigma}$ and $D+if$ and the fermions $\bar{\chi}_{-}$ and 
$\chi_{+}$. The corresponding twisted anti-chiral superfield,  
$\bar{\Sigma}$, obeys the constraints $\bar{D}_{-}\bar{\Sigma}
=D_{+}\bar{\Sigma}=0$.
\paragraph{}
Invariant Lagrangians for twisted chiral superfields $\Lambda_{A}$ can be 
constructed in direct analogy with those for conventional chiral superfields. 
The D-term Lagrangian (\ref{Dterm}) for chiral superfields 
is defined as an integral over all four of the fermionic 
coordinates. As the superspace measure $d^{4}\theta$ 
is invariant (up to an overall sign) 
under the interchange of $\theta_{+}$ and $\bar{\theta}_{+}$, an invariant
D-term for twisted chiral superfields can be constructed in exactly the same way,    
\begin{equation}
\tilde{\cal L}_{D}=\int\,d^{4}\theta\, {\cal K}(\Lambda_{A},\bar{\Lambda}_{A})
\label{tDterm}
\end{equation} 
where ${\cal K}$, is a real function of $\Lambda_{A}$ and $\bar{\Lambda}_{A}$. 
On the other hand the F-term Lagrangian (\ref{Fterm}) 
needs to be modified for twisted chiral superfields by interchanging the   
integrations over $\theta_{-}$ and $\bar{\theta}_{+}$.  
Thus we have a twisted F-term, 
\begin{equation}
\tilde{\cal L}_{F}=\int\,d^{2}{\vartheta} 
{\cal W}(\Lambda_{A}) \, \, + \int\, 
d^{2}\bar{{\vartheta}}\, \bar{{\cal W}}(\bar{\Lambda}_{A})
\label{tfterm}
\end{equation}
where the integration measures are 
$d^{2}{\vartheta}=d\theta_{-}d\bar{\theta}_{+}/2$ and 
$d^{2}\bar{{\vartheta}}=d\bar{\theta}_{-}d\theta_{+}/2$. Note also that $d^{4}\theta=
-d^{2}{\vartheta}d^{2}\bar{{\vartheta}}$
The twisted superpotential ${\cal W}$ is holomorphic in the 
twisted chiral superfields $\Lambda_{A}$. Both the D-term and the twisted F-term are 
invariant under $U(1)_{R}$, the F-term violates $U(1)_{A}$ unless the twisted 
superpotential has charge $+2$ under this symmetry. An example of a twisted superpotential 
which obeys this condition arises for the field strength superfield $\Sigma$. The 
Fayet-Iliopoulos $D$-term and the topological $\theta$-term for the gauge field multiplet 
can be combined in the form \cite{W1},  
\begin{eqnarray}
\tilde{\cal L}_{F} &= &\, \frac{i}{2}\int\,d^{2}{\vartheta}\, 
\tau\Sigma \,- \, \frac{i}{2} \int\,
d^{2}\bar{{\vartheta}}\, \bar{\tau}\bar{\Sigma} \nonumber \\
& =& \,-rD+\frac{\theta}{2\pi}f 
\label{f2}
\end{eqnarray}    
where the FI coupling $r$ and the vacuum angle $\theta$ combine to form a complex 
coupling $\tau=ir+\theta/2\pi$. 
The resulting F-term Lagrangian is 
a special case of the general expression (\ref{tfterm}) with twisted 
superpotential ${\cal W}=i\tau\Sigma/2$.  
\paragraph{}
Several general features of the vacuum structure and spectrum of effective theories 
which contain only twisted chiral superfields will be important in the following. 
Consider a theory of $M$ twisted chiral superfields $\Lambda_{A}$ $A=1,2\ldots,M$ 
with a low-energy effective Lagrangian 
of the form $\tilde{\cal L}=\tilde{\cal L}_{D}+ \tilde{\cal L}_{F}$. In particular, 
we will consider the case of a generic superpotential which breaks $U(1)_{A}$ symmetry.   
The bosonic terms in the Lagrangian are, 
\begin{equation}
\tilde{\cal L}_{\rm Bose}= g_{A\bar{B}}\left(-\partial_{\mu}\lambda^{A}
\partial^{\mu}\bar{\lambda}^{B}+E^{A}\bar{E}^{B}\right)+\left(
E^{A}\frac{\partial{\cal W}}{\partial\lambda^{A}} + \bar{E}^{A}
\frac{\partial\bar{{\cal W}}}{\partial\bar{\lambda}^{A}}\right)    
\end{equation}
With the conventions chosen above, the \Ka\ metric for twisted superfields is given by, 
\begin{equation} 
g_{A\bar{B}}=-\frac{\partial^{2} {\cal K}}{\partial \lambda^{A}\partial\bar{\lambda}^{B}} 
\label{kmetric}
\end{equation} 
After eliminating the auxiliary fields $E_{A}$ the bosonic Lagrangian becomes, 
\begin{equation}
\tilde{{\cal L}}_{\rm Bose}=-g_{A\bar{B}}\partial_{\mu}\lambda^{A}
\partial^{\mu}\bar{\lambda}^{B}-g^{A\bar{B}}
\frac{\partial{\cal W}}{\partial\lambda^{A}}
\frac{\partial\bar{{\cal W}}}{\partial\bar{\lambda}^{B}}    
\label{elim}
\end{equation} 
where $g^{A\bar{B}}$ is the inverse \Ka\ metric, 
$g_{A\bar{C}}g^{\bar{C}B}=\delta_{A}^{\,\,B}$.
\paragraph{}
The condition for a supersymmetric vacuum is the vanishing of the potential energy, 
\begin{equation} 
U=g^{A\bar{B}}
\frac{\partial{\cal W}}{\partial\lambda^{A}}
\frac{\partial\bar{{\cal W}}}{\partial\bar{\lambda}^{B}}=0     
\label{svac}
\end{equation}
In a unitary theory coordinates may be chosen so that $g^{A\bar{B}}$ is positive 
definite and thus the vacuum condition becomes $\partial {\cal W}/\partial \lambda_{A}=0$ 
for $A=1,2,\ldots,M$. This provides $M$ complex equations for the vacuum values of 
$\lambda_{A}$ which constitute $M$ complex unknowns. For a generic twisted 
superpotential ${\cal W}$, these conditions will have a finite number of 
isolated solutions. 
\paragraph{}
In two dimensions topologically stable solitons or kinks 
potentially occur in any theory with two 
or more isolated vacuum states. These are static 
solutions of the classical field equations 
which asymptote to two different 
vacua at left and right spatial infinity. Specifically we consider time-independent 
field configurations with boundary conditions $\lambda_{A}\rightarrow\alpha_{A}$ as 
$x\rightarrow -\infty$ and  $\lambda_{A}\rightarrow\beta_{A}$ as $x\rightarrow +\infty$ 
where $\alpha_{A}$ and $\beta_{A}$ are two different solutions of the vacuum condition 
(\ref{svac}). The mass of such a configuration obeys the following inequality \cite{FMVW} 
which hold for any complex constant $\gamma$ with $|\gamma|=1$, 
\begin{eqnarray}
M&= & \int_{-\infty}^{+\infty}\,dx\, \left[ g_{A\bar{B}}
\frac{\partial\lambda^{A}}{\partial x }
\frac{\partial\bar{\lambda}^{B}}{\partial x} + g^{A\bar{B}}
\frac{\partial{\cal W}}{\partial\lambda^{A}}
\frac{\partial\bar{{\cal W}}}{\partial\bar{\lambda}^{B}} \right] \nonumber \\
& = & \int_{-\infty}^{+\infty}\,dx\,\left|\frac{\partial\lambda^{A}}{\partial x }-
\gamma g^{A\bar{B}}\frac{\partial\bar{{\cal W}}}
{\partial\bar{\lambda}^{B}}\right|^{2} + 
\int_{-\infty}^{+\infty}\,dx\,\left[\bar{\gamma}
\frac{\partial{\cal W}}{\partial\lambda^{A}}
\frac{\partial\lambda^{A}}{\partial x }+\gamma
\frac{\partial\bar{{\cal W}}}{\partial\bar{\lambda}^{A}}
\frac{\partial\bar{\lambda}^{A}}{\partial x }\right] \nonumber \\ 
&\geq & \qquad{}  2{\rm Re}
\left[\bar{\gamma}\left({\cal W}(\beta^{A})-{\cal W}(\alpha^{A})\right)\right]
\label{bog1} 
\end{eqnarray}
By choosing $\gamma=\Delta{\cal W}/|\Delta{\cal W}|$ with $\Delta{\cal W}=
{\cal W}(\beta^{A})-{\cal W}(\alpha^{A})$ we obtain the Bogomol'nyi bound 
$M\geq 2|\Delta{\cal W}|$. 
\paragraph{}
As usual the Bogomol'nyi bound corresponds to a 
non-zero value for the central charge $Z$ appearing in the supersymmetry algebra 
(\ref{susyalg}). Thus we have $Z=2\Delta{\cal W}$.
This is consistent with our choice of a twisted F-term which breaks 
the $U(1)_{A}$ symmetry. In fact, this formula for the central charge is 
generic to all ${\cal N}=(2,2)$ supersymmetric theories \cite{CV2}, even those 
which do not have an effective Lagrangian of the simple Landau-Ginzburg form 
$\tilde{\cal L}_{D}+\tilde{\cal L}_{F}$. In all cases, it is possible to 
define an effective twisted superpotential ${\cal W}_{\rm eff}$ which has 
critical points at each vacuum and contributes to the central charge as 
$Z=\Delta{\cal W}_{\rm eff}$.  However, as explained in \cite{hnh}, 
there may also be additional contributions to the central charge from the generators of 
unbroken abelian global symmetries.   
\paragraph{}  
From (\ref{bog1}) we deduce that a BPS saturated 
soliton solution satisfies the first order equation, 
\begin{equation}
\frac{\partial\lambda^{A}}{\partial x }=
\frac{\Delta{\cal W}}{|\Delta{\cal W}|} g^{A\bar{B}}\frac{\partial\bar{{\cal W}}}
{\partial\bar{\lambda}^{B}}
\label{bogeq}
\end{equation}
A useful property of such BPS saturated 
solutions can be deduced by multiplying both sides of this 
equation by $\partial{\cal W}/\partial\lambda^{A}$, 
\begin{equation}
\frac{\partial{\cal W}}{\partial x}=  
\frac{\partial{\cal W}}{\partial\lambda^{A}}
\frac{\partial\lambda^{A}}{\partial x }=
\frac{\Delta{\cal W}}{|\Delta{\cal W}|} g^{A\bar{B}}
\frac{\partial{\cal W}}{\partial\lambda^{A}}
\frac{\partial\bar{{\cal W}}}{\partial\bar{\lambda}^{B}}=
\frac{\Delta{\cal W}}{|\Delta{\cal W}|}U({\cal W})
\label{straight}
\end{equation}
As the potential energy $U$ is real, this implies that a BPS soliton configuration 
corresponds to a straight line segment in the complex ${\cal W}$ plane joining the points 
${\cal W}(\alpha^{A})$ and ${\cal W}(\beta^{A})$. From the Bogomol'nyi bound, 
the mass of this soliton is $2|\Delta{\cal W}|$ which is the proportional to the 
length of the line segment. 
\paragraph{}

\section{The Classical Theory}
\paragraph{}
As in Section 1, we will consider a superrenormalizable  
$U(1)$ gauge theory with gauge superfield $V$ and  
field strength $\Sigma$ which is a twisted chiral superfield. 
The theory considered will also contain $N$ chiral superfields 
$\Phi_{i}$, $i=1,2,\ldots, N$ each of charge $+1$. A theory with only 
this chiral matter content would have a non-cancelling gauge anomaly in 
four-dimensions. However, in two dimensions the requirements for 
gauge anomaly cancellation are much less strict and the given matter content 
yields a consistent theory.     
The kinetic terms and minimal couplings for each of these fields can be 
written as a D-term in ${\cal N}=2$ superspace,
\begin{equation}
{\cal L}_{D}=\int\,d^{4}\theta\, \left[\sum_{i=1}^{N}\,\bar{\Phi}_{i}\exp(2V)\Phi_{i}
-\frac{1}{4e^{2}}\bar{\Sigma}\Sigma\right] 
\label{d1}
\end{equation}
The $U(1)$ gauge coupling $e$ has the dimensions of a mass, which means that 
the kinetic terms for the gauge multiplet are irrelevant in the infra-red \cite{W1}.  
As reviewed in the previous section, the Fayet-Iliopoulis term and and 
topological $\theta$-term can be combined in the twisted F-term, 
\begin{equation}
{\cal L}_{F}=\int\,d^{2}{\vartheta}\,  
{\cal W}(\Sigma) \, \, + \int\, 
d^{2}\bar{{\vartheta}}\, \bar{{\cal W}}(\bar{\Sigma})
\label{tfterm2}
\end{equation}
with the twisted superpotential ${\cal W}=i\tau\Sigma/2$. 
In two dimensions, the complexified coupling $\tau=ir+\theta/2\pi$ is 
dimensionless and corresponds to a 
marginal operator. In fact we will see that 
$\tau$ behaves very much like the complexified coupling 
$\tau_{4D}=i4\pi/g_{4D}^{2}+\theta_{4D}/2\pi$ in four-dimensional gauge theory.  
Both the F-term and the D-term are invariant 
under the full R-symmetry group, $U(1)_{A}\times U(1)_{R}$, at the classical level. 
The theory also has a global $SU(N)$ symmetry which acts on the flavour 
index $i$. In the following we will introduce ${\cal N}=(2,2)$ 
supersymmetric mass terms for the chiral multiplets, $\Phi_{i}$. 
However, we first consider the vacuum structure of the massless theory. 
\paragraph{}
Eliminating the auxiliary fields by their equations of motion, the classical 
potential energy is 
\begin{equation}
U=\sum_{i=1}^{N}|\sigma|^{2}|\phi_{i}|^{2} + e^{2}
\left(\sum_{i=1}^{N}|\phi_{i}|^{2}-r\right)^{2}
\label{cpotential}
\end{equation}
For a supersymmetric vacuum both terms in $U$ must vanish. For $r>0$, the 
vanishing of the second term requires that at least one $\phi_{i}$ must be 
non-zero. The vanishing of the first term then requires that $\sigma=0$. 
On the other hand, if $r=0$, then we must have $\phi_{i}=0$ for 
$i=1,2,\ldots N$ and then $\sigma$ is not 
constrained. In the former case the $U(1)$ gauge invariance is 
spontaneously broken 
while in the latter case it is preserved. For this reason we will refer 
to these two sets of vacua as the classical Higgs branch and classical 
Coulomb branch respectively 
although the usual idea of a moduli space of inequivalent vacua does not 
apply in two dimensions. If $r<0$, then there is no solution to the 
condition $U=0$ and, at least at the classical 
level, there are no supersymmetric vacua.        
\paragraph{}
The classical Higgs branch consists of the solution space of the equation, 
\begin{equation} 
\sum_{i=1}^{N}|\phi_{i}|^{2}=r
\label{const}
\end{equation}
modulo $U(1)$ gauge transformation which rotate each $\phi_{i}$ by the same 
phase; $\phi_{i}\rightarrow\exp(i\alpha)\phi_{i}$. This is precisely the 
definition of the complex projective space $CP^{N-1}$. The gauge degrees of 
freedom and the modes of the chiral fields which are orthogonal to the 
vacuum manifold acquire mass $\sqrt{r}e$ by the Higgs mechanism. In the 
low-energy limit $e\rightarrow\infty$, the kinetic term for the gauge multiplet   
vanishes, and the correponding component fields can be eliminated by their equations of motion. 
The resulting effective theory for the modes tangent to the vacuum manifold is 
an ${\cal N}=(2,2)$ supersymmetric 
$\sigma$-model with target space $CP^{N-1}$. 
The $CP^{N-1}$ target 
is covered by $N$ overlapping coordinate patches, 
${\cal P}_{j}$ with $j=0,1,\ldots,N$, each with 
$N-1$ gauge-invariant complex coordinates $w^{(j)}_{i}=\phi_{i}/\phi_{j}$ with $i\neq j$.  
The patch ${\cal P}_{j}$ covers the complement in $CP^{N-1}$ of the submanifold defined by 
$\phi_{j}=0$. In this patch, an unconstrained form of the $\sigma$-model action can 
be obtained by using the constraint (\ref{const}) to eliminate $\phi_{j}$.  
The superspace Lagrangian can be written in terms of chiral superfields, 
$W^{(j)}_{i}$, whose scalar components are $w_{i}^{(j)}$, 
\begin{equation}
{\cal L}_{\rm eff}=r \int\,d^{4}\theta\, 
\log\left(1+\sum_{i=1}^{N}\, '\,\bar{W}^{(j)}_{i}W^{(j)}_{i}\right)
\label{d3}
\end{equation} 
where $\sum '$ indicates that the term with $i=j$ in the sum is omitted. 
The superspace integrand is precisely the \Ka\ potential for the 
Fubini-Study metric on $CP^{N-1}$.  
Note that, in choosing particular a coordinate patch on the 
target space, we have concealed the $SU(N)$ symmetry of the model.  
The FI coupling of the underlying gauge theory is related to the 
dimensionless  coupling constant of the $\sigma$-model as $r=2/g^{2}$. 
Thus the low-energy $\sigma$-model is weakly coupled for $r>>1$. The 
FI parameter $r$ can also be thought of as the radius of the target space.
\paragraph{}
We will now consider how this 
analysis is modified in the presence of explicit mass terms for the 
chiral fields. The conventional supersymmetric 
mass term for chiral multiplets, which is familiar from four-dimensions,  
has the form of a superpotential bilinear in chiral superfields.  
In the present case, each of the chiral multiplets has the same charge 
and a mass term of this kind would violate gauge invariance. However, as pointed out 
recently by Hanany and Hori \cite{hnh}, there is another way of introducing 
a mass term which has no analog in four dimensions. 
In two dimensions the ${\cal N}=(2,2)$ gauge multiplet contains a 
complex scalar $\sigma$. On the classical Coulomb branch, for 
$\sigma\neq 0$, the potential (\ref{cpotential}) 
indicates that the chiral fields each acquire a mass $|\sigma|$ by the 
Higgs mechanism. Thus one may introduce a supersymmetric mass term 
for the chiral fields by coupling them to a background gauge multiplet 
in which the scalar field is frozen to its vacuum expectation value. 
As in \cite{hnh}, these will be refered to as twisted mass term.  
The new parameters can be introduced in a manifestly 
supersymmetric way by including a new gauge multiplet $\hat{V}_{i}$ 
for each twisted mass and constraining the corresponding field strengths  
$\hat{\Sigma}_{i}$ to the constant values $m_{i}$ by integrating over Lagrange multipliers 
which are themselves twisted chiral superfields.
\paragraph{} 
The resulting D-term Lagrangian for $N$  
chiral superfields of equal $U(1)_{G}$ charge with twisted masses $m_{i}$ is,  
\begin{equation}
{\cal L}_{D}=\int\,d^{4}\theta\, 
\left[\sum_{i=1}^{N}\,\bar{\Phi}_{i}\exp(2V+2\langle\hat{V}_{i}\rangle)\Phi_{i}
-\frac{1}{4e^{2}}\bar{\Sigma}\Sigma\right] 
\label{dd}
\end{equation}   
In the four-dimensional superspace notation of Section 1, 
the constant gauge background superfield, $\langle \hat{V}_{i} \rangle$, is given by, 
\begin{equation} 
\langle \hat{V}_{i} \rangle = -\theta^{\alpha}\sigma^{m}_{\alpha\dot{\alpha}}
\bar{\theta}^{\dot{\alpha}}\hat{V}_{mi}
\label{background}
\end{equation}
where $\hat{V}_{1i}={\rm Re}(m_{i})$, $\hat{V}_{2i}=-{\rm Im}(m_{i})$ and 
$\hat{V}_{0i}=\hat{V}_{3i}=0$. Note that only the differences between twisted masses 
are physically significant, the sum of the twisted masses can be absorbed by a constant 
shift in the complex scalar $\sigma$. Thus, without loss of generality, we may set 
$\sum_{i=1}^{N}m_{i}=0$.     
The introduction of twisted mass terms affects the classical symmetries of the model.  
As the twisted masses are the scalar components of a background 
gauge multiplet, they 
carry $U(1)_{A}$ charge $+2$. Hence, in the presence of twisted masses, 
$U(1)_{A}$ is explicitly broken to $Z_{2}$ while $U(1)_{R}$ remains unbroken.  
Generic twisted masses with $m_{i}\neq m_{j}$ also break 
the global $SU(N)$ flavour symmetry of the massless theory down to its maximal 
abelian subgroup $U(1)^{N-1}=(\otimes_{j=1}^{N} U(1)_{j})/U(1)_{G}$. 
As in Section 1, the $\Phi_{i}$ have charges $\delta_{ij}$ under $U(1)_{j}$      
\paragraph{}
In the presence of twisted masses, the potential energy becomes, 
\begin{equation}
U=\sum_{i=1}^{N}|\sigma+m_{i}|^{2}|\phi_{i}|^{2} + e^{2}
\left(\sum_{i=1}^{N}|\phi_{i}|^{2}-r\right)^{2}
\label{cpotential2}
\end{equation}
The vacuum structure for the Higgs branch with $r>0$ is now changed as 
follows. As before at least one chiral field, say $\phi_{k}$, must be 
non-zero for the second term to vanish. Now, for the first term to vanish 
also, we must have $\sigma=-m_{k}$. However, 
if we have generic twisted masses with $m_{i}\neq m_{j}$, this also requires 
$\phi_{i}=0$, for all $i\neq k$. 
Exactly one chiral field is non-zero in a given supersymmetric vacuum. 
It follows that there are 
exactly $N$ supersymmetric vacua ${\cal V}_{k}$ labelled by $k=1,2,\ldots,N$. 
Thus one effect of the introduction of twisted masses is to lift the 
classical vacuum moduli space, $CP^{N-1}$, leaving $N$ isolated vacua. 
As will be discussed below, this is consistent with the Witten index \cite{W3} 
for the $CP^{N-1}$ $\sigma$-model which is equal to $N$.   
In the vacuum ${\cal V}_{k}$ the scalar fields take 
values $\sigma=-m_{k}$ and $|\phi_{i}|=\sqrt{r}\delta_{ik}$. Hence there is exactly 
one vacuum ${\cal V}_{k}$ in each coordinate 
patch ${\cal P}_{k}$
\paragraph{}
As in the massless case, 
the low-energy effective action is obtained by taking the limit 
$e\rightarrow \infty$ and eliminating the gauge multiplet fields 
by their equations of motion. In the theory with twisted masses, this leads to a 
classically massive variant of the $CP^{N-1}$ $\sigma$-model. 
To describe the theory in the vacuum  
${\cal V}_{j}$, we choose the coordinate patch ${\cal P}_{j}$ with coordinates 
$w^{(j)}_{i}$ for $i\neq j$. The effective superspace Lagrangian becomes,     
\begin{equation}
{\cal L}_{\rm eff}=r \int\,d^{4}\theta\, 
\log\left(1+\sum_{i=1}^{N}\,'\,\bar{W}^{(k)}_{i}
\exp \left(2\langle\hat{V}_{i}\rangle-2\langle\hat{V}_{j}
\rangle\right) W^{(k)}_{i}\right)
\label{d4}
\end{equation}    
where, as in (\ref{d3}), $W^{(j)}_{i}$ are chiral superfields with scalar 
components $w^{(j)}_{i}$. The background twisted superfields 
$\langle \hat{V}_{i} \rangle $ are defined in (\ref{background}) above. 
\paragraph{}
To illustrate the properties of the effective theory with twisted masses, 
it will be useful to consider the simplest non-trivial case, $N=2$. 
In this case the low-energy theory involves a single chiral superfield, $W$, 
whose lowest component is the complex scalar field $w=\phi_{1}/\phi_{2}$. 
The component fields of 
$W$ are minimally coupled to a background gauge field 
$\langle\hat{V}_{1}\rangle-\langle \hat{V}_{2}\rangle$ 
whose scalar component is $m=m_{1}-m_{2}$. An explicit expression for the
low-energy effective Lagrangian in terms of component fields is given in 
Appendix A. The bosonic part of the effective Lagrangian is, 
\begin{equation}
{\cal L}^{(0)} =  
-\frac{1}{\rho^{2}}\left[r\left(\partial_{\mu}\bar{w}\partial^{\mu}w 
+|m|^{2}|w|^{2}\right)+\frac{\theta}{2\pi i}\varepsilon^{\mu\nu} 
\partial_{\mu}\bar{w}\partial_{\nu}w\right] 
\label{bosonic}
\end{equation}
where $\rho=1+|w|^{2}$. For $m=0$, (\ref{bosonic}) reduces to the 
standard Lagrangian of the $CP^{1}$ $\sigma$-model. 
For $m\neq 0$, the classical theory has two supersymmetric vacua 
which are located at $w=0$ and $w=\infty$ respectively. 
An equivalent Lagrangian which is 
non-singular in the neighbourhood of $w=\infty$ can be obtained by the coordinate 
transformation $w\rightarrow 1/w$. 

\section{The Classical BPS Spectrum}
\paragraph{}            
As described in Section 2, BPS states obey the 
exact mass formula $M=|Z|$ where $Z$ is the central charge 
appearing in the ${\cal N}=(2,2)$ supersymmetry algebra (\ref{susyalg}). 
In ${\cal N}=(2,2)$ theories the central charge recieves a contribution 
from the twisted F-terms in the Lagrangian. Specifically the contribution 
involves the difference of the twisted superpotential evaluated at 
left and right spatial infinity, denoted $\Delta{\cal W}$.  As in Section 
2, this corresponds to a topological charge carried by Bogomoln'yi saturated 
solitons. However, in theories which have unbroken abelian global symmetries the 
corresponding generators can also contribute to the central charge. 
In the classical theory described in the previous section the global $SU(N)$ 
symmetry is broken by the twisted masses $m_{i}$
to the abelian subgroup generated by the charges $S_{i}$. The corresponding formula 
for the central charge proposed in \cite{hnh} is, 
\begin{equation}
Z=2\Delta{\cal W}+i\sum_{i=1}^{N}m_{i}S_{i}
\label{ccharge}
\end{equation}
In the following, this formula will be checked explicitly for the 
{\em classical} theory whose low-energy description is the massive 
version (\ref{d4}) of the supersymmetric $CP^{N-1}$ $\sigma$-model. 
In particular, we will find that the classical theory has three different 
kinds of states which obey the BPS mass formula $M=|Z|$: 
those that carry only the global charges $S_{i}$, 
those that carry only topological charges $T_{i}$, and states which 
carry both kinds of charge. In this section we consider these three types of 
states in turn. In the subsequent sections, the relevance of this 
classical spectrum to the corresponding quantum theory will be explained. 
\subsubsection*{Elementary Quanta}
\paragraph{}
The fields appearing in the massive $\sigma$-model 
Lagrangian (\ref{d4}) transform under the global symmetries $U(1)_{i}$ and their 
quanta therefore carry the corresponding charges. In particular, consider 
the $N=2$ case described by (\ref{bosonic}) above.  Expanding 
around the vacuum at $w=0$, the tree-level spectrum includes a spinless 
particle of mass $M=|m|=|m_{1}-m_{2}|$ 
corresponding to the complex scalar field $w=\phi_{1}/\phi_{2}$. 
This scalar also has fermionic superpartner of the same mass.   
Both these states also carry $U(1)$ charges $S_{1}=-S_{2}=+1$, 
and therfore satisfy the BPS mass relation 
\begin{equation}
M=|m_{1}-m_{2}|=|S_{1}m_{1}+S_{2}m_{2}|=|Z|
\label{bps}
\end{equation}
Similarly, in the other supersymmetric vacuum $w=\infty$, the tree-level 
spectrum includes a BPS saturated multiplet whose scalar component has 
the quantum numbers of $w^{-1}=\phi_{2}/\phi_{1}$. 
\paragraph{}
The above analysis can easily be generalized for arbitrary $N$. 
In this case the theory has $N$ supersymmetric vacua 
${\cal V}_{i}$ $i=1,2,\ldots, N$. In the vacuum ${\cal V}_{j}$, the low-energy 
theory is described by the superspace Lagrangian (\ref{d4}) which contains 
$N-1$ chiral multiplets $W_{i}^{(j)}$ with $i\neq j$, 
described by the superspace Lagrangian (\ref{d4}). 
The particles corresponding to these fields have charge $S_{i}=+1$ and 
$S_{j}=-1$ and $S_{k}=0$ for $k\neq i,j$. Expanding the fields around their values 
in the vacuum ${\cal V}_{j}$ one 
finds that the component fields of $W^{(j)}_{i}$ have tree-level masses $M_{ij}=|m_{i}-m_{j}|$. 
Hence the corresponding states satisfy the BPS mass relation,   
\begin{equation}
M_{ij}=|m_{i}-m_{j}|=\left|\sum_{i=1}^{N}m_{i}S_{i}\right|=|Z|
\label{treelevel}
\end{equation}
\paragraph{}
In each of the $N$ supersymmetric vacuum states,  
the tree-level spectrum includes $N-1$ BPS saturated multiplets 
(and their anti-particles). Including the BPS states from each vacuum, 
there is one BPS multiplet corresponding to each of the $N(N-1)$ 
gauge-invariant fields $W^{(j)}_{i}=\Phi_{i}/\Phi_{j}$. 
In the next Section we will
consider quantum corrections to the classical BPS spectrum. One 
of the main results will be that 
the low-energy effective theory is weakly coupled for 
large values of the twisted masses. Hence, at least in this regime,  
the $N(N-1)$ multiplets found above 
are present in the BPS spectrum of the quantum theory. 
\subsubsection*{Solitons}
\paragraph{}
In general a two dimensional theory with isolated vacua can have topologically 
stable solitons or kinks. These are solutions of the classical equations of motion 
which interpolate between two different vacua at 
left and right spatial infinity. In the present case, where there are $N$ 
isolated vacua ${\cal V}_{i}$, it is useful to 
define $N$ topological charges $T_{i}$ as follows. 
A soliton which asymptotes to ${\cal V}_{k}$ at  $x=-\infty$ and 
${\cal V}_{l}$ at $x=+\infty$ has charges $T_{k}=-1$, $T_{l}=+1$ and $T_{i}=0$ for 
$i\neq k,l$. 
According to (\ref{ccharge}), the BPS mass formula for this soliton is, 
\begin{eqnarray}
M_{kl} & = & 2|\Delta {\cal W}| = 2\left|\sum_{i=1}^{N} {\cal W}(\sigma=-m_{i})T_{i}\right| 
\label{bps1}
\end{eqnarray}
where the twisted 
superpotential is given by its classical value, ${\cal W}=i\tau\sigma/2$. Thus 
\begin{eqnarray}  
M_{kl} & = & 2|{\cal W}(\sigma=-m_{l})-{\cal W}(\sigma=-m_{k})| 
\nonumber \\
& = & |m_{k}-m_{l}|\sqrt{r^{2}+\left(\frac{\theta}{2\pi}\right)^{2}}
\label{spectrum}
\end{eqnarray}
This suggests that, for generic twisted masses, the classical theory includes BPS solitons 
with $N(N-1)/2$ different masses. 
\paragraph{} 
It is straightforward to find explicit forms for these  soliton solutions. 
In the $N=2$ case, this may be accomplished by performing a change of variables in  
the bosonic Lagrangian (\ref{bosonic}). Specifically it is convenient to  
decompose the complex field $w$ in terms of its modulus and argument as, 
\begin{equation}
w=\tan\frac{\varphi}{2}\exp(i\alpha)
\label{cov2}
\end{equation}
where, in order to make the mapping one-to-one, we make the identifications 
$\varphi\sim\varphi+2\pi$ and $\alpha\sim\alpha+2\pi$.   
In terms of the new variables, the bosonic Lagrangian reads,  
\begin{equation}
{\cal L}_{\rm Bose}= -\frac{r}{4}\left[(\partial_{\mu}\varphi)^{2}+
\sin^{2}\varphi\left(|m|^{2}-(\partial_{\mu}\alpha)^{2}\right)\right] +
\frac{\theta}{4\pi}\epsilon^{\mu\nu}\partial_{\mu}(\cos\varphi)\partial_{\nu}\alpha
\label{sg}
\end{equation}
In this form, the bosonic sector of the model  
is a close relative of the sine-Gordon (SG) scalar field theory. 
In the present case there is an additional massless field $\alpha$ 
with derivative couplings to the SG field $\varphi$. Very similar, 
but not identical, 
generalizations of the SG theory have been considered in the literature 
before \cite{CSG}. The 
two SUSY vacua found above correpond to the two sets of zeros of the 
SG potential, which occur at $\varphi =2n\pi$ and at $\varphi=(2n+1)\pi$ 
for integer $n$. As $\alpha$ appears only through its derivatives it can 
take any constant value in the vacuum.       
\paragraph{}
The soliton solutions we seek can now be found explicitly by a trivial 
modification of the standard analysis of the sine-Gordon theory. 
The mass functional for time-dependent 
field configurations obeys the following inequalities, 
\begin{eqnarray}
M & = & \frac{r}{4}\int_{-\infty}^{+\infty}
dx\, \left(\frac{\partial\varphi}{\partial x}\right)^{2}
+\sin^{2}\varphi\left[ \left(\frac{\partial\alpha}{\partial x}\right)^{2}+|m|^{2}\right] 
\nonumber \\
& \geq & \frac{r}{4}\int_{-\infty}^{+\infty}\,
dx  \left(\frac{\partial\varphi}{\partial x}\pm |m|\sin\varphi\right)^{2}\mp 
2|m|\sin\varphi\frac{\partial\varphi}{\partial x} \nonumber \\
& \geq &\left| \frac{r}{4}\int_{-\infty}^{+\infty}\,
dx 2|m|\sin\varphi\frac{\partial\varphi}{\partial x}\right|\,\,=\,\, 
\left|\frac{rm}{2}
\left[\cos\varphi\right]_{-\infty}^{+\infty}\right| 
\end{eqnarray}
Hence defining the topological charge, $T$, 
\begin{equation}
T= -\frac{1}{2}\left[\cos\varphi\right]_{-\infty}^{+\infty}
\label{topcharge}
\end{equation}
we have the Bogomol'nyi bound $M\geq r|m||T|$. 
Configurations which saturate the bound 
must have $\alpha=$constant and $\varphi$ a solution of the sine-Gordon equation, 
\begin{equation}
\frac{\partial\varphi}{\partial x}= \pm |m|\sin\varphi
\label{sgeqn}
\end{equation}
In particular, the soliton which interpolates between two neighbouring 
vacua has $T=T_{1}=-T_{2}=1$ and is given by, 
\begin{equation}
\varphi=\varphi_{S}(x;|m|)=2\tan^{-1}\left[{\rm e}^{|m|x}\right]
\label{kink}
\end{equation}
where, for later convenience, we have introduced a notation which emphasizes 
the parametric dependence of the kink solution $\phi_{S}$ on 
$|m|$. The corresponding solution for $w$ has the simple form 
$w=w_{S}=\exp(|m|x)$. In addition to 
the usual degeneracy associated with spatial translations, the $T=1$ solution 
has an extra one-parameter degeneracy corresponding to the constant value 
of $\alpha$. This reflects the fact that the soliton solution breaks the $U(1)$ 
global symmetry generated by $S=(S_{1}-S_{2})/2$.      
The mass of the solution is given by the 
Bogomol'nyi formula; $M=r|m|T=r|m|$. As $m=m_{1}-m_{2}$, this agrees with the 
BPS mass formula (\ref{spectrum}) in the case $\theta=0$. However, the 
analysis given above is not altered in the presence of a non-zero vacuum angle and thus  
does not agree with the BPS mass formula (\ref{spectrum}) for $\theta\neq 0$. 
In fact the correct $\theta$ dependence of the soliton mass will be recovered 
only after a more precise semiclassical analysis given below 
which includes the possibility  of time-dependent classical solutions. 
\paragraph{}
Finding the $N(N-1)$ BPS saturated soliton solutions for the general case 
$N>2$ is now easy. The $N=2$ static solution found above 
can trivially be embedded in the appropriate $CP^{1}$ subspace of $CP^{N-1}$. 
Explicitly, for a soliton which interpolates between the vacua ${\cal V}_{k}$ and 
${\cal V}_{l}$ at left and right spatial infinity, we set $\phi_{i}=0$ for 
$i\neq k,l$ and the bosonic Lagrangian reduces to the $N=2$ expression 
(\ref{bosonic}) with 
$w=\phi_{k}/\phi_{l}$ and $m=|m_{k}-m_{l}|$. This yields a soliton solution 
of mass $M_{kl}=r|m_{k}-m_{l}|$. Again this agrees with the BPS mass formula in the 
case $\theta=0$.  
\paragraph{}
\subsubsection*{Dyons}
\paragraph{}
In the previous section we discovered that the soliton solution of the 
low-energy effective theory which interpolates between the $k$'th and the $l$'th vacuum 
has a degeneracy associated with global $U(1)$ rotations generated by $S=(S_{l}-S_{k})/2$. 
The soliton therefore has a periodic collective coordinate, $\alpha\in [0,2\pi]$ 
which is analogous to the $U(1)$ `charge-angle' of a BPS monopole in four-dimensions. 
In the four-dimensional case \cite{TW}, allowing the collective coordinate to 
become time-dependent yields dyonic solutions 
which carry both magnetic and electric charges. This suggests that we should 
look for solutions of the equations of motion with time-dependent collective coordinate, 
$\alpha(t)$, which carry global $U(1)$ charge in addition to their topological charge. 
In the case $N=2$ the corresponding conserved charge is determined by 
Noether's theorem to be, 
\begin{equation}
S=\frac{r}{2}\int_{-\infty}^{+\infty}dx\,\sin^{2}\varphi\,\dot{\alpha} + 
\frac{\theta}{4\pi}\int_{-\infty}^{+\infty}dx\, \partial_{x}(\cos\varphi)
\label{scharge}
\end{equation}
The first term confirms that solutions with non-zero $\dot{\alpha}$ will generically 
have non-zero charge $S$. The signifigance of the second term will be explained below. 
\paragraph{}
For the $N=2$ case, finding the desired time-dependent solutions is almost 
trivial: setting $\alpha=\omega t$ in the classical equation of motion 
simply has the effect of shifting the mass parameter $|m|^{2}\rightarrow |m|^{2}-\omega^{2}$. 
Hence, in the notation introduced in (\ref{kink}) above, the solution for $\varphi$ is, 
\begin{equation}
\varphi=\varphi_{\omega}=\varphi_{S}(x;\sqrt{|m|^{2}-\omega^{2}})        
\label{risky}
\end{equation}
or, equivalently $w=\exp(\sqrt{|m|^{2}-\omega^{2}}\,x+i\omega t)$. 
Setting $\theta=0$ for the moment, the mass and charge of this 
solution may be evaluated to give,
\begin{eqnarray}
M= \frac{r|m|^{2}}{\sqrt{|m|^{2}-\omega^{2}}}    
& \qquad{} \qquad{} \qquad{} &  S=\frac{r\omega}{\sqrt{|m|^{2}-\omega^{2}}}    
\label{mns}
\end{eqnarray}
Evidently the solution only makes sense for $\omega<|m|$. Physically this correspond to 
the threshold above which the rotating soliton is unstable to the 
emission of soft quanta of the charged field.   
\paragraph{} 
Classically the angular velocity $\omega$ and therefore the charge $S$ can take any 
value. In the quantum theory, the elementary quanta of the field $w$ 
carry charge $+1$ and thus we 
expect that $S$ is quantized in integer units. 
In the weak-coupling limit of the quantum theory we perform a 
semiclassical quantization of the dyon system. 
In this case, the allowed values of the charge are determined by 
the Bohr-Sommerfeld quantization condition. For a time-dependent solution of 
period $\tau=2\pi/\omega$, the action-per-period,  
${\cal S}_{\tau}$, is quantized according to the 
condition,      
\begin{equation}
{\cal S}_{\tau}+M\tau=2\pi n
\label{bs}
\end{equation}
where $n$ is an integer
The action-per-period of the dyon solution $\phi=\phi_{\omega}$, $\alpha=\omega t$ 
can be evaluated to obtain, 
\begin{equation}
{\cal S}_{\tau}+M\tau=\frac{2\pi r\omega}{\sqrt{|m|^{2}-\omega^{2}}}=2\pi S
\label{app}
\end{equation}
Hence the Bohr-Sommerfeld quantization condition (\ref{bs}) implies that the allowed values 
of $S$ are integers as expected. The semiclassical spectrum of the model includes 
an infinite tower of dyon states with topological charge $T=1$ 
labelled by an integer $U(1)$ charge $S$ with masses, $M=|m|\sqrt{S^{2}+r^{2}}$. 
These two-dimensional dyons are close relatives of the Q-kinks studied in \cite{QK}. 
\paragraph{}
It is straightforward to repeat the argument for non-zero vacuum angle $\theta$. 
Both the $U(1)$ charge of the periodic solution and its the action-per-period 
are shifted by the contribution of the final term in the Lagrangian (\ref{sg}) 
\begin{eqnarray}
S\rightarrow S-\frac{\theta}{2\pi} & \qquad{} \qquad{} \qquad{} & {\cal S}_{\tau}
\rightarrow  {\cal S}_{\tau}-\frac{\theta}{2\pi}
\label{shift}
\end{eqnarray} 
while the mass and period of the solution are unchanged. For 
non-zero $\theta$, even the static soliton solution with $\omega=0$ acquires a 
non-zero $U(1)_{S}$ charge. The net effect of the two 
shifts appearing  in (\ref{shift}) above cancels in (\ref{bs}) and the Bohr-Sommerfeld 
condition again implies that the charge 
$S$ is quantized in integer units. The mass spectrum of the dyons becomes, 
\begin{equation}
M=|m|\sqrt{\left(S+\frac{\theta}{2\pi}\right)^{2}+r^{2}}
\label{sat2}
\end{equation}
This result is an exact analog of the Witten effect \cite{WE} 
for dyons in four-dimensional gauge theories. 
In particular the state with $S=0$ has precisely the $\theta$-dependent 
mass predicted in (\ref{spectrum}) above. 
As $\theta$ increases from $0$ to $2\pi$ the charge of each state is shifted 
by one-unit; $S \rightarrow S+1$. As states with each integer value of $S$ 
are present in the theory, the spectrum is invariant under this shift.         
\paragraph{}
The classical spectrum of the $N=2$ model consists of particle states labelled by two 
seperate 
charges: the global $U(1)$ charge $S$ and the topological charge $T$.  We 
can now deduce a simple mass formula which applies to all the states we have 
discussed so far: the elementary particles and anti-particles with $(S,T)$ charges 
$(\pm 1,0)$, the soliton and dyon states with charges $(S,1)$ and the 
charge conjugate states with 
charges $(-S,-1)$. 
\begin{equation}
M_{S,T}=|m|\sqrt{\left(S+\frac{\theta}{2\pi}T\right)^{2}+r^{2}T^{2}}
\label{sat3}   
\end{equation} 
This is consistent with the BPS mass formula $M=|Z|$ with the central charge,  
\begin{equation}
Z=-mi\left(S+\tau T \right)
\label{ccharge4}
\end{equation}
where $\tau$ is the complexified coupling constant $ir+\theta/2\pi$. 
This analysis is easily extended to the case of general $N$. 
The soliton which interpolates between the vacua  ${\cal V}_{k}$ and ${\cal V}_{l}$ 
carries topological charge $T=(T_{l}-T_{k})/2=+1$ and has a collective 
coordinate associated with global $U(1)$ rotations generated by $S=(S_{l}-S_{k})/2$. 
After applying the Bohr-Sommerfeld quantization condition  (\ref{bs}) we find an 
infinite tower of dyons with integral $U(1)$ charges. The dyon masses are given by 
(\ref{sat3}) with the replacement $m\rightarrow m_{l}-m_{k}$
\paragraph{}
To summarize the results of this section, we have found that the classical 
BPS spectrum includes both elementary particles 
which carry the global $U(1)$ Noether charges $S_{i}$ and solitons which interpolate 
between different vacua and carry topological charges $T_{i}$. 
The two sets of charges can be written as two $N$-component 
vectors $\vec{S}=(S_{1},S_{2},\ldots S_{N})$ and 
$\vec{T}=(T_{1},T_{2},\ldots T_{N})$. The spectrum 
includes all states with one non-zero charge vector (which can be either $\vec{S}$ or $\vec{T}$) 
of the form $\pm (0,\ldots,+1, \ldots,-1,\ldots,0)$. For each (non-zero) allowed value of the 
topological charge $\vec{T}$, the spectrum also includes an infinite tower of dyons with 
global charge vector $\vec{S}=S\vec{T}$ where $S$ can be any integer.  
The masses of all these states obey the BPS mass formula $M=|Z|$ with, 
\begin{equation}
Z=-i\vec{m}\cdot(\vec{S}+\tau\vec{T}) 
\label{ccharge2}
\end{equation}
where $\vec{m}=(m_{1},m_{2},\ldots,m_{N})$. 
\paragraph{}
The spectrum described above bears a striking resemblance to the classical spectrum 
of BPS states of an ${\cal N}=2$ supersymmetric Yang-Mills theory in four dimensions 
with gauge group $SU(N)$ (see, for example, the introduction of \cite{DS}). 
The four-dimensional theory has a dimensionless gauge 
coupling $g_{4D}$ and vacuum angle $\theta_{4D}$ which can be combined in a single 
complex coupling $\tau_{4D}=4\pi/g^{2}_{4D}+i\theta_{4D}/2\pi$. The theory 
contains a complex scalar field $A$ in the adjoint representation of the gauge group  
The vacuum expectation value of this field is an $N\times N$ 
matrix with complex eigenvalues 
$a_{i}$ which obey $\sum_{i=1}^{N}a_{i}=0$. For generic non-zero $a_{i}$, $SU(N)$ is broken 
to its maximal abelian subgroup $U(1)^{N-1}$ and the 
the gauge bosons corresponding to the $N(N-1)$ broken generators get masses 
by the Higgs mechanism. These states and their superpartners are 
charged under the $U(1)$ 
factors of the unbroken gauge group. The charges can be represented as vectors  
$\vec{q}=(q_{1},q_{2},\ldots,q_{N})$ with integer entries $q_{i}$ obeying 
$\sum_{i=1}^{N}q_{i}=0$. There are two massive gauge bosons for each of the 
$N(N-1)$ possible charge vectors of the form 
$\vec{q}=\pm (0,\ldots,+1, \ldots,-1,\ldots,0)$   
The theory 
also contains BPS monopoles which carry the corresponding magnetic charges 
$\vec{h}=(h_{1},h_{2},\ldots,h_{N})$ for integers $h_{i}$ with $\sum_{i=1}^{N}h_{i}=0$.  
For each $U(1)$ the unit of magnetic charge is determined by the Dirac quantization condition to be 
$4\pi/g_{4D}^{2}$. In these units, the allowed magnetic charge vectors 
also have the form $\vec{h}=\pm (0,\ldots,+1, \ldots,-1,\ldots,0)$.     
Finally, for each allowed charge vector $\vec{h}$, the theory has 
an infinite tower of dyons which carry magnetic charge $\vec{h}$ 
and electric charge $\vec{q}=Q\vec{h}$, for each integer $Q$.  
Each of these states gives 
rise to a BPS multiplet of ${\cal N}=2$ supersymmetry with mass 
$M=\sqrt{2}|{\cal Z}|$ where, 
\begin{equation}
{\cal Z}=\vec{a}\cdot(\vec{q}+\tau_{4D}\vec{h})   
\label{4dspectrum}
\end{equation}
and $\vec{a}=(a_{1},a_{2},\ldots,a_{N})$. 
\paragraph{}
The correspondence between the two theories is such that 
the global $SU(N)$ symmetry of the 
$CP^{N-1}$ $\sigma$-model corresponds to the $SU(N)$ gauge symmetry of the four-dimensional theory. 
The Noether charges $\vec{S}$ corresponds to the electric charge vector $\vec{q}$ 
while the topological 
charges $\vec{T}$ is the analog of the magnetic charge $\vec{h}$. Up to an overall normalization, 
the twisted masses $\vec{m}$ correspond to the vacuum expectation value $\vec{a}$ 
and the dimensionless coupling $\tau$ 
is identified with the four-dimensional coupling $\tau_{4D}$. Each BPS multiplet in the 
two-dimensional theory has a counterpart in the four-dimensional theory. Note, however, that 
while the ${\cal N}=2$ theory in four dimensions has $N-1$ massless photons, the two-dimensional 
theory has no massless particles. Also note that, so far, the correspondence between 
the two theories is strictly classical. In Section 6, we will find a modified 
version of the correspondence which applies at the quantum level.           
\section{Quantum Effects}
\paragraph{}
In this section we will consider how quantum corrections affect the 
analysis of the low-energy theory given above. We start with the case of vanishing 
twisted masses, $m_{i}=0$ 
where the low-energy effective theory for $r>0$ is the supersymmetric $CP^{N-1}$ 
$\sigma$-model (\ref{d3}). The $\sigma$ model coupling $g$ is related to the FI 
parameter as $g=\sqrt{2/r}$. 
In the weak-coupling regime $r>>1$, one may apply 
perturbation theory in the $\sigma$-model coupling. At one loop, one finds 
a logarithmic divergence which can be removed by renormalizing the bare 
coupling constant $g_{0}$ as \cite{FR,SVZ}, 
\begin{equation}
\frac{1}{g^{2}(\mu)}=
\frac{1}{g^{2}_{0}}-\frac{N}{8\pi}\log\left(\frac{M_{\rm UV}^{2}}{\mu^{2}}\right)
\label{renorm}      
\end{equation}
where $g(\mu)$ is the renormalized coupling constant. 
Here $M_{UV}$ is an ultra-violet regulator (for example a Pauli-Villars mass 
as in \cite{SVZ})  and $\mu$ is the RG subtraction point. 
It is shown in \cite{FR} that this is the 
only renormalization required to all orders in perturbation theory. The 
form of the one-loop correction (\ref{renorm}) tells us that the model is 
asymptotically free. The converse to this fact is that the $\sigma$-model flows to 
strong coupling in the IR and hence 
perturbation theory is of no use in calculating low energy quantities such 
as the masses of particles. In particular perturbation theory breaks down 
at energy scales $\mu$ of the order of the RG invariant scale, 
\begin{equation}
\Lambda=\mu\exp\left(-\frac{4\pi}{Ng^{2}(\mu)} \right)
\label{lam}
\end{equation}
As mentioned above, the true spectrum of the model bears 
no relation to the field content of the $CP^{N-1}$ $\sigma$-model 
Lagrangian (\ref{d3}). While the classical theory contains massless 
degrees of freedom 
corresponding to gauge-invariant fields of the form $w_{i}=\phi_{i}/\phi_{j}$, 
the quantum theory has a mass-gap. For example, in the case $N=2$ 
the true spectrum consists of two massive particles which form a doublet of 
the  $SU(2)$ flavour symmetry and there are no asymptotic states 
with the quantum numbers of the field $w=\phi_{1}/\phi_{2}$ \cite{W2}.   
\paragraph{}
Another quantum effect which appears at one-loop in the massless theory is an anomaly 
in the $U(1)_{A}$ symmetry. The divergence of the $U(1)_{A}$ current has a non-vanishing 
contribution from a one-loop `diangle' diagram. The resulting violation of $U(1)_{A}$ 
charge is, 
\begin{equation}
\Delta Q_{A}=\frac{N}{\pi}\int\, d^{2}x \, f_{\rm E}=2Nk
\label{anom}
\end{equation}
where $f_{\rm E}$ is the Euclidean field strength of the gauge field and 
$k$ is the topological charge. Thus $U(1)_{A}$ is explicitly broken to $Z_{2N}$ by 
the anomaly. As usual 
the anomaly means that the vacuum angle $\theta$ can be 
set to zero by a $U(1)_{A}$ rotation of the fields and is no longer a physical 
parameter of the theory. Both the 2D gauge theory and the 
$\sigma$-model to which it reduces at low energy, 
have instantons with integer topological charge. The anomaly formula 
(\ref{anom}) reflects the fact that the $k$-instanton solution of 
the $CP^{N-1}$ $\sigma$-model has a total of $2Nk$ fermion zero modes. Each of these 
zero modes carry one unit of $U(1)_{A}$ charge. Two 
zero modes correspond to the action of the supersymmetry generators which act 
non-trivially on the instanton solution and two more correspond to the action of 
superconformal generators. The remaining zero modes do not correspond to symmetries. 
\paragraph{}
In the light of the classical correspondence discussed in the previous section,  
it is interesting to compare the two quantum effects described above with the quantum 
behaviour of the corresponding ${\cal N}=2$ theory in four dimensions. 
In the four-dimensional theory, the gauge coupling runs logarithmically at one loop. 
However, the coefficient of the logarithm which appears in the four-dimensional 
version of (\ref{renorm}) is exactly twice that of the two-dimensional theory. 
In addition, the four-dimensional theory also has an anomalous $U(1)$ R-symmetry. 
A charge-$k$ $SU(N)$ Yang-Mills instanton in the minimal ${\cal N}=2$ theory has 
$4Nk$ fermion zero modes which is twice the number of the corresponding $\sigma$-model 
instanton. Hence, the violation of $R$-charge in the background of a four-dimensional 
instanton is twice that given in (\ref{anom}). Equivalently 
the four-dimensional anomaly breaks the $U(1)$ R-symmetry down 
to $Z_{4Nk}$ compared to the residual $Z_{2Nk}$ of the two-dimensional theory. 
In fact, both these discrepancies can be resolved by considering a four-dimensional 
theory with $N$ additional hypermultiplets rather than the minimal ${\cal N}=2$ theory. 
In this case the renormalization of the $\sigma$-model coupling is precisely the 
same as that of the gauge coupling. The residual discrete R-symmetries of the two 
theories also agree. This correspondence will be refined below.     
\paragraph{}
In the case of non-zero twisted masses $m_{i}\neq 0$, 
the analysis of the quantum theory is somewhat different. 
As described above, the low-energy theory is now a variant of the 
$CP^{N-1}$ $\sigma$-model with an explicit mass term for 
each of the scalar fields $w_{i}$ (and superpartners) which 
appear in the classical action (\ref{d4}). 
As in the massless case, a divergent one-loop 
diagram leads the logarithmic renormalization of the $\sigma$-model 
coupling given in (\ref{renorm}).   
However, now the particles running around the loop have masses $|m_{i}-m_{j}|$. 
As before we will start with the simplest case $N=2$ where there 
is a single massive charged particle with mass $|m|=|m_{1}-m_{2}|$. For 
energy scales much less than $|m|$ the massive particle decouples and the $\sigma$-model 
coupling is frozen at a fixed value $g_{\rm eff}\simeq g(\mu=|m|)$.  
This result, which will be confirmed in an explicit one-loop calculation 
in the soliton background below, means that the low-energy theory is weakly coupled 
for $|m|>>\Lambda$. For $N>2$, the various particles appearing in the tree-level 
spectrum can have different masses and the story is more complicated; 
the heaviest particle decouples at the highest energy scale leaving an effective 
theory with one less flavour and so on. However the fact remains that if all the 
bare masses $|m_{i}-m_{j}|$  are much larger than the RG-invariant scale 
$\Lambda$, then $g_{\rm eff}<<1$ and the low-energy theory is weakly coupled. 
Thus we come to the opposite conclusion to the one reached above for the 
massless case: perturbation theory applied to the massive $\sigma$-model 
action (\ref{d4}) should be reliable. 
\paragraph{}
According to the above discussion of the case $N=2$, low-energy quantities such as the 
classical soliton mass, $M_{0}=r|m|$, which depend on 
the FI parameter at the classical level should be corrected at one loop by the replacement, 
\begin{equation}  
r\rightarrow r_{\rm eff}(|m|)=\frac{1}{2\pi}\log\left(\frac{|m|^{2}}{\Lambda^{2}}\right)
\label{replace}
\end{equation}
As this effect is an important ingredient in understanding the BPS spectrum of the model an 
explicit derivation of this result is given below. 
In any theory, the one-loop correction to the mass of a soliton 
comes from performing a Gaussian path-integral over the fluctuations of the fields 
around their classical values. Explicitly the one-loop correction is given by\cite{DHN}, 
\begin{equation}
M_{1}=M_{0}+\frac{\hbar}{2}\left(\sum \lambda_{\rm B} - \sum \lambda_{\rm F}\right) + 
\Delta M_{\rm ct}
\label{oneloop}
\end{equation}
where $\lambda_{\rm B}$ and $\lambda_{\rm F}$ are the frequencies 
of the normal modes for small fluctuations of the bose and 
fermi fields respectively. 
$\Delta M_{\rm ct}$ is the contribution of the one-loop 
counter-term, which appears in the renormalization (\ref{renorm}) of the $\sigma$-model 
coupling, evaluated on the soliton background. Explicitly, for $N=2$, 
the Lagrangian counter-term corresponding to (\ref{renorm}) is, 
\begin{equation}
\Delta{\cal L}_{\rm ct}=\frac{\delta r}{r}{\cal L} =\frac{1}{\pi r}\log
\left(\frac{M_{\rm UV}}{\mu}\right)\,{\cal L}
\label{lct1}
\end{equation} 
Thus the contiribution of the counterterm to the one-loop soliton mass 
is
\begin{equation}
\Delta M_{\rm ct}=\int^{+\infty}_{-\infty}\, dx 
\Delta {\cal L}_{\rm ct}[\varphi_{S}(x)] = \frac{\delta r}{r}M_{\rm cl}
\label{fkg}
\end{equation}
As the counter-term contribution is divergent, the only way to get 
a finite answer for the soliton mass is if the sum appearing in (\ref{oneloop}) 
is also divergent and the two divergences cancel.  
\paragraph{}
In order to determine the frequencies $\lambda_{\rm B}$ and 
$\lambda_{\rm F}$, we need to expand the Lagrangian 
to quadratic order in each of the fields around the 
classical solution $\varphi_{\rm cl}=\varphi_{S}(x)$, 
$\alpha_{\rm cl}=0$. The fluctuating fields can be written as two real scalar 
fields $\vec{u}^{\rm T}=(u_{1},u_{2})$ and two Majorana fermions 
$\vec{\rho}_{\alpha}^{\rm T}=(\rho_{1\alpha },\rho_{2\alpha})$ , 
with $\vec{\rho}^{\,\dagger}_{\alpha}=\vec{\rho}^{\,T}_{\alpha}$. As 
in Section 2, the index $\alpha$ runs over the two values $-$ and $+$.     
\begin{equation}
{\cal L}={\cal L}[\varphi_{\rm cl},\alpha_{\rm cl}]-\frac{r}{4}
\left(\vec{u}\cdot{\cal M}_{\rm B}\cdot\vec{u}+
\vec{{\rho}}^{\,T}_{\alpha}\cdot
{\cal M}_{\rm F}^{\alpha\beta}\cdot\vec{\rho}_{\beta}\right)
\label{expand}
\end{equation} 
With a judicious choice of basis for the fluctuating fields 
which is given explicitly in Appendix B, the 
differential operators ${\cal M}_{\rm B}$ and ${\cal M}_{\rm F}$, simplify to,  
\begin{eqnarray}
{\cal M}_{\rm B}^{ij}= \delta^{ij}\left(\frac{\partial^{2}}{\partial t^{2}}
+\Delta_{\rm B}\right) & \qquad{} \qquad{} &  {\cal M}^{ij}_{\rm F}=  
\delta^{ij}\left(\begin{array}{cc} \frac{\partial}{\partial t} & D \\ -D^{T} &
\frac{\partial}{\partial t}\end{array}\right) 
\label{expansion}
\end{eqnarray}
with 
\begin{eqnarray}
D=\frac{\partial}{\partial x}+m\cos\left(\varphi_{S}(x)\right) & \qquad{} & 
D^{T}=-\frac{\partial}{\partial x}+m\cos\left(\varphi_{S}(x)\right)
\label{ddt}
\end{eqnarray}
and 
\begin{equation}
\Delta_{\rm B}=DD^{T}= -\frac{\partial^{2}}{\partial x^{2}}+
|m|^{2}\cos\left(2\varphi_{S}(x)\right)
\label{delb}
\end{equation}
In this basis, the fluctuating degrees of freedom consists of 
two-decoupled copies of the fluctuations around the soliton of 
the supersymmetric sine-Gordon model and the analysis from this 
point on is an easy generalization of the calculation given 
by Kaul and Rajaraman \cite{KR}. Separating out the time dependence of 
the bosonic fluctuations as 
$\vec{u}=\vec{\mu}\exp(i\lambda_{\rm B}t)$ 
the resulting eigenvalue problem we need to solve is,  
\begin{equation}
\Delta_{\rm B}\vec{\mu}(x)=\lambda^{2}_{\rm B}\vec{\mu}(x)
\end{equation} 
\paragraph{}
The operator $\Delta_{\rm B}$ appears in the corresponding calculation 
in the sine-Gordon model and its spectrum is well known. In particular, it has 
a single normalizable zero mode $u^{(0)}(x)=\varphi'_{S}(x)/\sqrt{M}$. 
Hence the vector $\vec{\mu}$ has two linearly-independent 
zero modes; $(u^{(0)},0)$ and $(0,u^{(0)})$. As usual, the 
bosonic zero-modes should correspond to collective coordinates 
of the soliton. As in any two-dimensional theory 
the configuration $\varphi_{\rm cl},\alpha_{\rm cl}$ 
yields a one-parameter family of solutions under spatial translations  
$x\rightarrow x+X$. 
However, as explained in the previous section, 
there is also an additional degeneracy which is not present in the 
ordinary sine-Gordon model: the soliton has a periodic  
collective coordinate, $\alpha$, which corresponds to global $U(1)$ rotations. 
Thus the number of collective coordinates matches the number of bosonic zero modes. 
In addition to the zero modes $\Delta_{\rm B}$ has a continuous spectrum 
of scattering modes scattering eigenstates 
$u^{(k)}(x)$ with eigenvalues 
$\lambda_{\rm B}=\sqrt{k^{2}+|m|^{2}}$ and phase-shift, 
\begin{equation}
\delta_{\rm B}(k)=-2\tan^{-1}\left(\frac{k}{|m|}\right)
\label{pshift}
\end{equation} 
\paragraph{}
The fluctuations, $\vec{\rho}_{\alpha}$,  of the the fermi fields 
are governed by the equation ${\cal M_{F}}^{\alpha\beta}\cdot
\vec{\rho}_{\beta}=0$. From (\ref{expansion}), we see that 
this equation has time-dependent solutions of the form;
\begin{equation}
\vec{\rho}_{\pm}=\vec{\alpha}_{\pm}
\exp\left(-i\lambda_{\rm F}t\right)+ \vec{\alpha}_{\pm}^{\,*}
\exp\left(+i\lambda_{\rm F}t\right)
\label{ftdep}
\end{equation}
where $\vec{\alpha}_{\pm}$ satisfy the equations, 
\begin{eqnarray}
iD\vec{\alpha}_{+}=
-\lambda_{\rm F}\vec{\alpha}_{-} &  \qquad{} &
-iD^{T}\vec{\alpha}_{-}=-\lambda_{\rm F}\vec{\alpha}_{+}
\label{feqs1}
\end{eqnarray}
Acting on the first and second equation with $-iD^{T}$ and $iD$ respectively 
we find, 
\begin{eqnarray}
D^{T}D\vec{\alpha}_{+}=
\Delta_{\rm B}\vec{\alpha}_{+}=\lambda^{2}_{\rm F}\vec{\alpha}_{+}
&  & DD^{T}\vec{\alpha}_{-}= \left(-\frac{\partial^{2}}
{\partial x^{2}}+|m|^{2}\right)\vec{\alpha}_{-}=\lambda^{2}_{\rm F}
\vec{\alpha}_{-}
\label{feqs2}
\end{eqnarray}
The spinor component $\vec{\alpha}_{+}$ obeys the same equation as the bosonic 
fluctuation $\vec{\mu}$. This component therefore has two normalizable zero 
modes and a continuum of scattering states labelled by the momentum $k$ 
with $\lambda_{\rm F}=
\sqrt{k^{2}+|m|^{2}}$. As above the scattering for momentum $k$ occurs with 
a phase shift $\delta^{+}_{\rm F}$ given by (\ref{pshift}). 
On the other hand, the other spinor 
component $\vec{\alpha}_{-}$ is an eigenstate of 
the free massive Klein-Gordon operator. 
This operator has no normalizable zero modes. It does have scattering 
eigenstates with $\lambda_{\rm F}=\sqrt{k^{2}+|m|^{2}}$ although, because there 
is no interaction, the corresponding phase shift is zero, $\delta^{+}_{\rm F}=0$.
\paragraph{}
Taken together, the above facts imply that the fermion fields have a net total 
of two normalizable zero modes; $\vec{\alpha}_{+}=(\varphi'_{S}/\sqrt{M},0)$ and 
$\vec{\alpha}_{+}=(0,\varphi'_{S}/\sqrt{M})$ with $\vec{\alpha}_{-}=0$ in both 
cases. Like the bosonic zero modes, these modes can be interpreted 
in terms of the symmetries of the theory which are broke by the soliton 
solution. In particular, the fermion zero modes correspond to 
supersymmetry generators which act non-trivially on the 
soliton field configuration. The fact that there are only two zero modes is 
significant: it means that there are only two such generators. Conversely, the 
soliton solution must be invariant under two of the four supersymmetries of the 
theory. This confirms that the solitons of this model 
are BPS saturated. 
\paragraph{}
Turning to the non-zero modes, the above analysis indicates that the bose  
and fermi fields have the same non-zero eigenvalues: a continuum starting 
at $\lambda_{\rm F}=|m|$. Naively, this suggests that 
the contributions of bose and fermi fluctuations should cancel in   
the one-loop correction to the soliton mass (\ref{oneloop}). In fact, as 
explained in \cite{KR}, this reasoning is not correct. For a continuous 
spectrum of eigenvalues the sums appearing in (\ref{oneloop}) are replaced by 
integrals over a density of eigenvalues. Placing the system in a spatial box 
of length $L$, the density of states is given by the formula \cite{KR}; 
\begin{equation}
\rho(k)=\frac{d{\rm n}}{dk}=\frac{1}{2\pi}\left[L + \frac{d\delta}{dk}\right]
\label{density}
\end{equation}
As we saw above, one component of the fermion fluctuations has zero phase shift 
while the remaining component and the bose fluctuations both have the non-zero 
phase shift (\ref{pshift}). Thus, 
although the non-zero eigenvalues of the bose and 
fermi fields are equal, the densities of these eigenvalues are not. The sums 
appearing in (\ref{oneloop}) can be evaluated as; 
\begin{eqnarray}
\sum \lambda_{\rm B} - \sum \lambda_{\rm F}  & =& \int^{+\infty}_{-\infty} 
dk\, \sqrt{k^{2}+|m|^{2}}\left(\rho_{\rm B}(k) -\rho_{\rm F}(k)\right) \nonumber \\
&= & 2 \int^{+\infty}_{-\infty}\frac{dk}{2\pi}\,\sqrt{k^{2}+|m|^{2}}
\left[ \frac{d\delta_{\rm B}}{dk}-\frac{1}{2}
\left(\frac{d\delta^{+}_{\rm F}}{dk}+ \frac{d\delta^{-}_{\rm F}}{dk}\right)
\right] 
\nonumber \\
& =& -4|m|\int^{\infty}_{0}\frac{dk}{2\pi}\, \frac{1}{\sqrt{k^{2}+|m|^{2}}}
\label{bfsums}
\end{eqnarray}
The final integral is logarithmically divergent. Introducing a UV cut-off 
$M_{UV}$ as the upper limit of this integral, the one-loop formula for the 
soliton mass becomes, 
\begin{equation}
M_{1}=M_{\rm cl}-\frac{m}{\pi}\log\left(\frac{M_{UV}}{|m|}\right) + 
\Delta M_{\rm ct}
\label{m1}
\end{equation}
From (\ref{fkg}) we have, 
\begin{equation}
M_{1}=\left(1+\frac{\delta r}{r}\right)M_{\rm cl}-
\frac{|m|}{\pi}\log\left(\frac{M_{UV}}{|m|}\right)=
|m|\left(r(\mu)-\frac{1}{\pi}\log\left(\frac{\mu}{|m|}\right)\right)
\end{equation}
The $\mu$ dependence of the soliton mass can be eliminated using 
the definition (\ref{lam}) 
of the RG invariant scale $\Lambda$. The final result is 
\begin{equation}
M_{1}=\frac{|m|}{\pi}\left[\log\left(\frac{|m|}{\Lambda}\right)+ C \right]   
\label{finalmass}
\end{equation}
As expected the classical and one-loop masses are related by the replacement 
(\ref{replace}) of the tree-level FI coupling by a low-energy effective FI coupling. 
In fact only the divergent parts of the one-loop correction were evaluated carefully 
in the above analysis. The undetermined numerical constant $C$ reflects the possible  
contribution of finite-counter terms at one loop.  
In particular, to evaluate $C$, 
one would have to carefully compare the renormalization prescriptions  
used in the vacuum and one-soliton sectors. For a recent treatment of this issue 
see\footnote{Note however, that the model considered here has twice as many supersymmetries 
as the minimal supersymmetric theories considered in \cite{WO,KR}. In particular, 
the above theory has short multiplets and therefore 
quantum corrections must respect the BPS mass formula. In the case of 
the minimal theories this conclusion can be avoided as pointed out by 
\cite{NSNR}.} \cite{NSNR}.   
\paragraph{}
The analysis of the $U(1)_{A}$ anomaly is also modified by the 
introduction of twisted masses. In the massless case, the $U(1)_{A}$ anomaly implied 
the vacuum angle, $\theta$, could be shifted by a $U(1)_{A}$ rotation of the fields. 
In this case theories with different values of the vacuum angle $\theta$ are 
physically equivalent. Non-zero twisted masses, explicitly break $U(1)_{A}$ down 
to $Z_{2}$. However, as twisted masses 
correspond to the expectation values of background twisted chiral superfields,  
the symmetry can be restored by assigning $U(1)_{A}$ charge $+2$ to the masses.  
Hence a simultaneous rotation of the fields and the twisted 
masses is required to shift the vacuum angle.  Specializing to the $N=2$ case 
considered above, the anomaly means that theories with different values of $\theta$ 
are equivalent to theories with the same value of $\theta$ which 
differ in the complex phase $\exp(i\beta)$ of the mass parameter $m$. Taking into 
account the coefficient of the anomaly (\ref{anom}), this means that 
physical quantities should depend on $\theta$ and $\beta$ in the combination    
$\theta_{\rm eff}=\theta-2\beta$. In the previous section we discovered a two-dimensional 
analog of the Witten effect for BPS dyons: the global $U(1)$ charge, $S$, of the 
dyon was shifted by an amount $\theta/2\pi$. According to the above discussion
we should replace $\theta$ by $\theta_{\rm eff}$ in the mass formula (\ref{sat2}). 
Thus the spectrum of BPS states undergoes a non-trivial monodromy 
as we follow a large circle in the complex $m$ plane. 
In particular, as $m\rightarrow \exp(2\pi i)m$ the spectrum of dyons transforms as; 
\begin{eqnarray}
(S,1)\rightarrow (S-2,1) & \qquad{} \qquad{} & (S,-1)\rightarrow (S+2,-1)
\label{monod}
\end{eqnarray}
As the elementary particle invariant under this transformation, the 
monodromy of a BPS state with charges $(S,T)$ can be compactly written as 
$(S,T)\rightarrow (S-2T,T)$.       
\paragraph{}
In fact it is 
straightforward to exhibit the above monodromy 
directly in the one-loop analysis of the 
quantum corrections to the soliton and dyon masses. 
As in the previous section we expand 
to quadratic order in fluctuations of the fields around the classical background. 
However, in this case the classical background is the time-dependent dyon solution,  
$\varphi_{\rm cl}=\varphi_{\omega}(x)$ and $\alpha_{\rm cl}=\omega t$. The resulting expansion 
of the Lagrangian is,  
\begin{equation}
{\cal L}={\cal L}[\varphi_{\rm cl},\alpha_{\rm cl}]-\frac{r}{4}
\left(\vec{u}\cdot{\cal M}_{\rm B}\cdot\vec{u}+
i\bar{\Psi}_{\alpha}
\hat{\cal M}_{\rm F}^{\alpha\beta}\Psi_{\beta}\right)
\label{expand2}
\end{equation}  
In this case we have chosen to parametrize the fermionic fluctuations in terms 
of a single Dirac fermion $\Psi$ rather than two Majorana fermions. The corresponding 
$\gamma$-matrices are $\gamma^{0}=i\sigma_{2}$, $\gamma^{1}=-\sigma_{1}$ and $\gamma^{5}=
\gamma^{0}\gamma^{1}=-\sigma_{3}$. Details of the calculation leading to (\ref{expand2}) 
are given in Appendix B. In this basis 
the differential operator $\hat{\cal M}_{\rm F}$ can be written as, 
\begin{eqnarray}
\hat{\cal M}_{\rm F} & = & 
\gamma^{\mu}\partial_{\mu}+\cos\left(\varphi_{\omega}(x)\right)
\left[i\omega\gamma^{0}+{\rm Re}(m)I+i{\rm Im}(m)\gamma^{5}\right] \nonumber \\ 
& = & \gamma^{\mu}(\partial_{\mu}+a_{\mu})+\cos\left(\varphi_{\omega}(x)\right)
|m|\exp\left(i\beta\gamma^{5}\right)
\label{fquadratic}
\end{eqnarray}
where, as above, $\beta={\rm arg}(m)$ and 
we have defined an auxiliary, two-dimensional gauge field $a_{\mu}$ 
which has components 
$a_{0}=\omega\cos\left(\varphi_{\omega}(x)\right)$ and $a_{1}=0$. 
Further details of the expansion in fluctuations around the soliton background  
are given in Appendix B. At one loop, the dependence of 
dyon masses on the phase $\beta$ 
can only come from the $\beta$ dependence in the formula (\ref{fquadratic}) for 
$\hat{\cal M}_{\rm F}$. The one-loop contribution of fermion fluctuations to the 
effective action in the soliton sector is, 
\begin{equation}
S_{\rm F}=\frac{1}{2}\log\det\left[\hat{\cal M}_{\rm F}\right]  
\label{seffm}
\end{equation}   
where, 
\begin{equation}  
{\det}^{\frac{1}{2}}\left[\hat{\cal M}_{\rm F}\right]=
\int {\cal D}\Psi{\cal D}\bar{\Psi}\exp\left[i\int\,d^{2}x\,
\bar{\Psi}\hat{\cal M}_{\rm F}\Psi \right]
\label{detm}
\end{equation}
The dependence of the exponent on the phase $\beta$ can be absorbed by a chiral 
transformation of the Dirac fields; 
\begin{eqnarray}
\Psi\rightarrow \Psi'=\exp\left(i\frac{\beta}{2}\gamma^{5}\right)\Psi 
& \qquad \qquad{} &  \bar{\Psi}\rightarrow \bar{\Psi}'=\bar{\Psi}
\exp\left(i\frac{\beta}{2}\gamma^{5}\right)
\label{psiprime}
\end{eqnarray}
which gives, 
\begin{equation}
\bar{\Psi}\hat{\cal M}_{\rm F}\Psi_{\beta}=\bar{\Psi}'
\hat{\cal M}^{(0)}_{\rm F}\Psi'_{\beta}
\end{equation}
with 
\begin{equation}
\hat{\cal M}^{(0)}_{\rm F}
 =  \gamma^{\mu}(\partial_{\mu}+a_{\mu})+|m|\cos\left(\varphi_{\omega}(x)\right)
\label{mf0}
\end{equation}
Naively it appears that by performing this change of variables in the path integral 
we can prove that the determinant appearing in (\ref{detm}) does not depend on 
$\beta$. However, this argument incorrectly assumes that the path integral 
measure is invariant under chiral rotations. The standard analysis of the chiral 
anomaly (see for example Chapter 9 of \cite{PS}) reveals that the measure transforms 
under (\ref{psiprime}) by a phase 
factor which depends on the topological charge of the background fields.   
For a two-dimensional Dirac fermion coupled 
to a background gauge-field $a_{\mu}$, the measure transforms 
with a non-trivial Jacobian: ${\cal D}\Psi{\cal D}\bar{\Psi}=
{\cal J}{\cal D}\Psi'{\cal D}\bar{\Psi}'$ where, 
\begin{eqnarray} 
{\cal J} & = & \exp\left[-i\beta
\int\,d^{2}x\,\frac{1}{2\pi}\varepsilon^{\mu\nu}\partial_{\mu}a_{\nu}\right]
\label{jac} 
\end{eqnarray} 
Using the explicit form of the gauge field $a_{\mu}$ given above we have,  
\begin{eqnarray}
& =& \exp\left[\frac{i\beta\omega}{2\pi}
\int\,d^{2}x\,\partial_{x}\left(\cos(\varphi_{S}(x))\right)\right]  
\label{measure}
\end{eqnarray}
The exponent of the Jacobian factor provides a one-loop correction to the 
action-per-period of the dyon solution, 
\begin{eqnarray}  
{\cal S}_{\tau} & = &
\int_{0}^{\tau}dt\int_{-\infty}^{+\infty}dx\,\left[{\cal L}[\varphi_{\rm cl},\alpha_{\rm cl}] + 
\frac{\beta\omega}{2\pi}\partial_{x}\left(\cos(\varphi_{S}(x))\right)\right] \nonumber \\
& =&  2\pi S-M\tau-\theta+ 2\beta T 
\label{measure2}
\end{eqnarray}
As $T=1$ for the dyon, we find that $\theta$ enters in the Bohr-Sommerfeld quantization condition 
in the combination ${\theta}_{\rm eff}(m)=\theta-2\beta$ as expected. This dependence 
leads to the non-trivial monodromy (\ref{monod}). 
\paragraph{}
In the previous section we found that the $N=2$ theory, 
has a spectrum of BPS states with charges $(S,T)$ which obey a mass 
formula $M=|Z|$ with, $Z=-im(S+\tau T)$. In this section 
we have exhibited one-loop effects which replace 
the parameters $r$ and $\theta$ which appear in the central charge 
by their effective counterparts $r_{\rm eff}$ defined 
in (\ref{replace}) and $\theta_{\rm eff}=\theta-2\arg(m)$. These two 
effects combine to replace $\tau$ with,   
\begin{equation}
\tau_{\rm eff}=ir_{\rm eff}+\theta_{\rm eff}/2\pi=
\frac{i}{\pi }\log\left(\frac{m}{\Lambda}\right)
\label{tau1loop}
\end{equation}
Importantly $\tau_{\rm eff}$, and thus the central charge itself, is a 
holomorphic function of $m$ and $\Lambda$. This holomorphic dependence 
could have been anticipated because both the twisted masses and the 
$\Lambda$-parameter correspond to the expectation values of background twisted 
chiral superfields. As discussed in Section 1, the resuting contribution to the central 
charge corresponds to the difference between the asymptotic values of 
some twisted superpotential ${\cal W}_{\rm eff}$.   
which is holomorphic function of the twisted chiral superfields. 
In the next section the relevant twisted superpotential will 
be determined exactly. 
\paragraph{}
So far we have determined the central charge to one-loop in the weak-coupling 
expansion of the low-energy $\sigma$-model (\ref{bosonic}). 
In the remainder of this section we will discuss the possible form of 
higher-order corrections. As discussed above, we expect that the exact central 
charge will have the form, $Z=-i(mS+m_{D}T)$ where 
$m_{D}=\Delta{\cal W}_{\rm eff}$ is a holomorphic 
function of $m$ and $\Lambda$ which is approximately equal to $\tau_{\rm eff}m$ 
for $|m|>>\Lambda$. The correspondence with ${\cal N}=2$ theories in 
four-dimensions suggests that the exact form of $m_{D}$ can be further constrained 
by the following arguments \cite{S}. First, by RG invariance, the BPS 
masses can only depend on the FI coupling through the dynamical scale $\Lambda$. 
Holomorphy and dimensional analysis then imply that 
perturbative corrections in the $\sigma$-model coupling correspond 
to inverse powers of $\log(m/\Lambda)$. However, 
the weak-coupling monodromy indicates that $m_{D}$ can only have 
the single branch-cut which arises at one-loop. 
The additional branch cuts 
introduced by higher-loop corrections would necessarily 
spoil the monodromy of the BPS spectrum 
discovered above. In particular, the spectrum of the theory 
would no longer be uniquely defined for given values of $m$ and $\Lambda$. 
Thus, for self-consistency, all the higher-order 
perturbative corrections to the one-loop formula for $m_{D}$ must vanish.  
\paragraph{}
Next we consider possible non-perturbative corrections to the one-loop 
result for $m_{D}$ given above. The $U(1)_{A}$ symmetry of the 
classical theory is restored if the twisted mass-parameter $m$ is assigned charge 
$+2$. As $m_{D}$ depends linearly on the effective twisted superpotential, it 
must also have charge $+2$ if the twisted F-term in the effective theory is to be 
$U(1)_{A}$ invariant. The classical result $m_{D}=\tau m$ is consistent with this 
charge assignement. In the quantum theory, $U(1)_{A}$ is explicitly broken to 
$Z_{4}$. The anomaly equation (\ref{anom}) implies that, in the sector of the 
theory with topological charge $k$, $U(1)_{A}$ is violated by $4k$ units. 
This corresponds to a correction to $m_{D}$ of the form $m(\Lambda/m)^{2k}$. 
Using the explicit form of the $\Lambda$-parameter (\ref{lam}), this 
correction depends on the $\sigma$-model coupling as $\exp(-4\pi k/g^{2})$. 
This is precisely the dependence expected for the leading semiclassical contribution of $k$ 
$\sigma$-model instantons.    
\paragraph{} 
To summarize the main results of this section, the BPS mass formula 
 the $N=2$ theory is $M=|Z|$ where $Z=-i(mS+m_{D}T)$ where $m_{D}$ is a holomorphic 
function of $m$ and $\Lambda$ with the weak coupling expansion, 
\begin{equation}
m_{D}=\frac{im}{\pi}\left[\log\left(\frac{m}{\Lambda}\right)+ 
\sum_{k=1}^{\infty}\,c_{k}\left(\frac{\Lambda}{m}\right)^{2k}\right]
\label{infser2}
\end{equation}
where $c_{k}$ are undetermined numerical coefficients. 
The first term in (\ref{infser2}) comes from one-loop perturbation 
theory while the term proportional to $c_{k}$ is a $k$-instanton contribution. 
In the next section we will find the exact formula for $m_{D}$ and 
thereby determine the unknown coefficients $c_{k}$. Note that the form 
of $(\ref{infser2})$ is consistent with the modified version of the classical 
correspondence described in the previous section: the 
quantum-corrected BPS spectrum implied by (\ref{infser2}) has precisely the same 
form as that of ${\cal N}=2$ supersymmetric gauge theory in four-dimensions 
with gauge group $SU(N)$ and $N$ additional hypermultiplets in the fundamental 
representation. Specifically the parameters $m$ and $m_{D}$ correspond to the 
two BPS charge vectors $a$ and $a_{D}$ in the Seiberg-Witten solution of the 
four-dimensional theory. In the classical correspondence the complex coupling 
$\tau$ was identified its four-dimensional counterpart; 
$\tau_{4D}=i4\pi/g_{4D}^{2}+\theta_{4D}/2\pi$. Similarly, at the quantum level, 
the $\Lambda$-parameters of the two theories are identified. 
This quantum correspondence will be made precise in the next section.           

\section{The Exact BPS Spectrum}
\paragraph{}
So far we have analysed the two-dimensional gauge theory 
introduced in Section 1 in a particular region of parameter space 
where it reduces to a massive variant of the supersymmetric 
$CP^{N-1}$ $\sigma$-model. This requires that the dimensionful gauge coupling 
$e$ is much larger than the dynamical scale $\Lambda$.  
The characteristic feature 
of this regime is that the $U(1)$ gauge symmetry is 
spontaneously broken and the gauge degrees of freedom get large masses 
from the Higgs mechanism. 
In this case the theory has  instantons which give 
a complicated series of corrections to the BPS mass spectrum. For the case 
$N=2$, the general form of these corrections, 
which involve contributions from all numbers of instantons, was given in (\ref{infser2}) 
above. On the other hand, as discussed in Section 1, 
the masses of BPS state are independent of the 
gauge-coupling and therefore we may attempt to calculate them in another 
more favourable region of parameter space. In the following, we will 
analyse the BPS spectrum in the case $e<<\Lambda$ 
where the theory has very different 
behaviour from that described above. We will find that 
the $U(1)$ gauge symmetry is unbroken and the matter multiplets, rather than 
the gauge multiplet, are very massive. In this regime, 
the weakly-coupled description of the theory does not have instantons and an 
exact calculation of the BPS spectrum is possible.  
\paragraph{}
As discussed above, the condition to decouple the gauge theory modes 
and obtain a $CP^{N-1}$ $\sigma$-model at low energies is $e>>\Lambda$.  
In this context it is useful to think of the mass scale $e$ as the effective 
UV cut-off for the low-energy $\sigma$-model. As in any asymptotically 
free theory, a large heirarchy between the cutoff and the 
$\Lambda$-parameter arises if the theory is weakly coupled at 
the cut-off scale. Thus the condition $e>>\Lambda$ can be written 
in terms of the renormalized $\sigma$ model coupling  
(\ref{renorm}) as $g^{2}(\mu=e)<<1$ or, in terms of the FI parameter, 
$r(\mu=e)>>0$. In the following, we want to consider the theory in the 
opposite limit $e<<\Lambda$, which according to (\ref{renorm}) corresponds 
to the regime of {\em negative} FI parameter, $r(\mu=e)<<0$. We begin by considering 
the model without twisted masses. The classical bosonic potential given in Section 3 was,        
\begin{equation}
U=\sum_{i=1}^{N}|\sigma|^{2}|\phi_{i}|^{2} + e^{2}
\left(\sum_{i=1}^{N}|\phi_{i}|^{2}-r\right)^{2}
\label{cpotential2B}
\end{equation}
For $r<0$, the states of minimum energy have $\phi_{i}=0$ with arbitrary 
$\sigma$. 
However, the energy density of these minima does not vanish as is required 
for a supersymmetric vacuum state, but has the non-zero value 
$e^{2}r^{2}$. Thus, at first sight, it appears that supersymmetry is 
spontaneously broken for $r<0$. On the other hand, the 
Witten index of the $r>0$ theory 
(or, equivalently, that of the $CP^{N-1}$ $\sigma$-model) is equal to $N$. 
By standard arguments \cite{W3}, this should 
preclude supersymmetry breaking for any value of $r$. 
\paragraph{}
The resolution of this apparant paradox was given by Witten in \cite{W1}: 
in order to obtain the correct vacuum structure for $r<0$, it is necessary to include 
quantum effects. In this case the minima of the potential (\ref{cpotential2B})   
generically have $\sigma\neq 0$ as described above, 
and therefore the chiral multiplet scalars, $\phi_{i}$, and 
their superpartners, gain masses equal to $|\sigma|$. 
In the region of field space $|\sigma|>>e$, these chiral multiplets become 
very massive and can be 
integrated out to give a low-energy effective action for the twisted 
superfield $\Sigma$ with scalar component $\sigma$. In the same region of field 
space, the gauge coupling is small compared to the other mass scales in the theory and 
perturbation theory in the gauge coupling is reliable. Note that in the previous sections 
we considered a different weak-coupling expansion, namely perturbation theory in the 
$\sigma$-model coupling. Gauge theories in two dimensions are super-renormalizable and, 
in the present case, the 
only divergent diagram appears at one-loop \cite{W1}. 
This logarithmic divergence can be removed by 
a renormalization of the FI parameter $r$, 
\begin{equation}
r(\mu)=r_{0}-\frac{N}{4\pi}\log\left(\frac{M_{UV}^{2}}{\mu^{2}}\right)
\label{renorm2}
\end{equation}
where $M_{UV}$ is a UV cut-off and $\mu$ is the RG subtraction scale. 
Notice that the form of this renormalization is identical to that 
which occured in the $r>0$ theory. In particular, just as in that case, 
we may eliminate the renormalized FI parameter $r(\mu)$ in favour of the 
RG invariant scale; 
\begin{equation} 
\Lambda=\mu\exp\left(-\frac{2\pi r(\mu)}{N} \right)
\label{lam2}
\end{equation}
\paragraph{}
Including terms with up to two derivatives or four fermions, 
the effective action has the form, 
\begin{equation} 
{\cal L}_{\rm eff}=\int  d^{4}\vartheta \,\, {\cal K}_{\rm eff}[\Sigma,\bar{\Sigma}] + 
\int d^{2}\vartheta\,\, {\cal W}_{\rm eff}[\Sigma] +
\int d^{2}\bar{\vartheta}\,\,\bar{{\cal W}}_{\rm eff}[\bar{\Sigma}]
\label{leff}
\end{equation}
After performing the renormalization of the FI parameter described above, 
the effective twisted superpotential is given at one loop by \cite{DDDS,W1}, 
\begin{equation}
{\cal W}_{\rm eff}=\frac{i}{2}\Sigma\left(\hat{\tau}-\frac{N}{2\pi i}
\log\left(\frac{2\Sigma}{\mu}\right)\right)
\label{weff}
\end{equation}
In fact, as the massive fields which are integrated out to obtain (\ref{weff}) only appear 
quadratically in the action, the one-loop result is actually exact (see Appendix A of 
\cite{CV2}). 
The complexified coupling constant 
$\hat{\tau}$ is equal to $ir(\mu)+\theta/2\pi+n^{*}$ where the integer $n^{*}$ 
is chosen to minimize the potential energy. As explained in \cite{W1}, this 
minimization of the potential reflects the fact 
that a non-zero value of the $\theta$ parameter in two-dimensions corresponds 
to a constant background electric field \cite{COL}. The states of the system 
with $n\neq n^{*}$, are unstable to pair creation of 
charged particles which screens the background field leaving the state with 
$N=n^{*}$. The corresponding potential energy is, 
\begin{equation}
U=g^{\Sigma\bar{\Sigma}}\left
|\frac{\partial{\cal W}_{\rm eff}}{\partial\sigma}\right|^{2}
\label{pe2}
\end{equation}
where $g^{\Sigma\bar{\Sigma}}=(g_{\Sigma\bar{\Sigma}})^{-1}$ is the 
inverse of the \Ka\ metric, 
\begin{equation}
g_{\Sigma\bar{\Sigma}}=-\frac{\partial^{2}{\tilde K}_{\rm eff}}
{\partial\Sigma\partial
\bar{\Sigma}}
\label{eeff} 
\end{equation}
\paragraph{}
Supersymmetric vacua correspond to the zeros of $U$. 
As the \Ka\ potential recieves only finite corrections of order 
$e^{2}/\sigma^{2}$, for large $\sigma$ we can be confident that the 
inverse metric $g^{\Sigma\bar{\Sigma}}$ is non-vanishing. Thus the relevant 
condition for a supersymmetric vacuum is the 
vanishing of $\partial{\cal W}_{\rm eff}/\partial\sigma$ which requires, 
\begin{equation}
\sigma^{N}=\left(\frac{\mu}{2e}\right)^{N}\exp\left(2\pi i \tau(\mu)\right)=
\tilde{\Lambda}^{N}
\label{vacuumeq}
\end{equation}
where $\tilde{\Lambda}=\Lambda\exp(-1+i\theta/N)/2$. 
Equation (\ref{vacuumeq}) has $N$ solutions,  
\begin{equation}
\sigma=\tilde{\Lambda}\exp\left(\frac{2\pi in}{N}\right)
\label{sigma1}
\end{equation} 
where $n=1,2,\ldots,N$. 
Thus the one-loop corrected effective theory has $N$ 
supersymmetric vacua even though the classical theory had none. 
This number coincides with the known value of the Witten index of the 
supersymmetric $CP^{N-1}$ $\sigma$-model. It also 
matches the number of classical vacua found above 
in the Higgs phase with non-zero twisted masses. 
As in the Higgs phase, the chiral anomaly 
breaks $U(1)_{A}$ down to $Z_{2N}$. Note however, that 
topological charge is not quantized in the Coulomb phase and 
the Euclidean field equations do not have instanton solutions \cite{W2}.  
Note also that the non-zero vacuum values for $\sigma$ spontaneously 
break the residual $Z_{2N}$ symmetry down to $Z_{2}$.         
\paragraph{}
The theory has $N$ isolated supersymmetric vacua and can therefore 
have BPS saturated solitons which interpolate between these vacua.  
As the low energy 
effective theory includes only twisted chiral superfields, we can immediately 
apply formulae given at the end of Section 2. According to (\ref{bog1}), 
a soliton solution interpolating between the vacua with $n=l$ and $n=k$ 
at left and right spatial infinity has mass $M=|Z|$ where, 
\begin{equation} 
Z=2\Delta{\cal W}_{\rm eff}= 2\left[ {\cal W}_{\rm eff}\left(\sigma=\exp\frac{2\pi il}{N}\right)- 
{\cal W}_{\rm eff}\left(\sigma=\exp\frac{2\pi ik}{N}\right)\right]
\label{lmmass}
\end{equation}
Because of the spontaneously broken $Z_{2N}$ symmetry, 
the resulting mass only depends on the difference $p=l-k$ and, 
for each $p$, is given by 
\begin{equation}
M_{k}=\frac{N}{\pi}\left|\exp\left(\frac{2\pi ip}{N}\right)-1\right|
\tilde{\Lambda} 
\label{m=0spectrum}
\end{equation}
one can also check that for $p=1,2\ldots N$ the degeneracy 
of BPS states is, \begin{equation}
D_{p}=\left(\begin{array}{c} N \\ 
                             p \end{array}\right) 
\label{degen}
\end{equation}   
The lightest soliton states with $p=1$ and degeneracy $N$ are interpreted as the 
elementary quanta of the the chiral fields $\Phi_{i}$ \cite{W2}. 
In fact they carry charge $+1$ under the unbroken $U(1)$ gauge symmetry.  
These states transform in the fundamental 
representation of the flavour symmetry group $SU(N)$. The states with $p>1$ correspond 
to stable bound states of $p$ different flavours of 
elementary quanta and therefore transform in the $p$'th antisymmetric tensor representation 
of $SU(N)$. The degeneracy $D_{p}$ in (\ref{degen}) is equal to the 
dimension of this representation and is therefore consistent with this 
interpretation. 
In contrast, the elementary quanta associated with the fields of the 
gauge multiplet are not BPS saturated \cite{W2}. 
\paragraph{}
The analysis given above depended on integrating out the chiral multiplets. 
This step is only valid if the masses of these fields are much larger 
than those of the gauge multiplet, which requires $|\sigma|>>e$. As 
the vacua in (\ref{sigma1}) are located at $\sigma\sim \Lambda$, the 
analysis is self-consistent for $e<<\Lambda$, as advertised in Section 1. 
In the opposite limit $e>>\Lambda$, 
the low-energy theory becomes the $CP^{N-1}$ $\sigma$-model and 
the true vacuum states have $\phi_{i}\neq 0$ and spontaneously broken 
$U(1)$ gauge symmetry. Thus the theory moves from a Coulomb phase 
to a Higgs phase as the gauge coupling is increased. 
Despite this, there are several quantities in the theory which are 
independent of $e$. The simplest of these is the Witten index itself which 
is equal to $N$ for all values of the parameters. The theory also has another 
supersymmetric index, introduced in \cite{CFIV}, 
which is independent of D-term parameters such as $e$ and depends 
holomorphically on twisted F-term parameters such as $\tau$ and $m_{i}$. 
Importantly, for the present purposes, these include the BPS mass spectrum. 
Thus the mass formula (\ref{m=0spectrum}) is exact for all values of $e$ and 
going to the regime $e>>\Lambda$, (\ref{m=0spectrum}) gives 
the exact BPS spectrum of the 
supersymmetric $CP^{N-1}$ $\sigma$-model. In particular, the 
mass spectrum agrees with that obtained by other methods 
which exploit the integrability of the model to obtain its exact 
spectrum and S-matrix \cite{KK,EH}.    
\paragraph{}
Only a small modification of this analysis is needed to include non-zero 
twisted masses \cite{hnh}. 
The same reasoning as before leads to an effective twisted superpotential; 
\begin{equation}
{\cal W}_{\rm eff}=\frac{i}{2}\left(\hat{\tau}\Sigma-{2\pi i}\sum_{i=1}^{N}(\Sigma+m_{i})
\log\left(\frac{2}{\mu}(\Sigma+m_{i})\right)\right)
\label{weff2}
\end{equation}
Setting $\partial{\cal W}_{\rm eff}/\partial\Sigma=0$, the resulting condition for a 
supersymmetric vacuum state can be written as, 
\begin{equation}
\prod_{i=1}^{N}(\sigma+m_{i})-\tilde{\Lambda}^{N}=\prod_{i=1}^{N}(\sigma-e_{i})=0
\label{vacuumeq2}
\end{equation} 
Thus there are $N$ supersymmetric vacua at $\sigma=e_{i}$ for $i=1,\dots,N$. 
Consider a soliton obeying the boundary conditions, 
$\sigma \rightarrow e_{k}$ as $x\rightarrow -\infty$ and 
$\sigma \rightarrow e_{l}$ as $x\rightarrow +\infty$. 
The soliton mass is given by $M_{kl}=|Z_{kl}|$ where 
a short calculation reveals that, 
\begin{eqnarray} 
Z_{kl}& = & 2\left[{\cal W}_{\rm eff}(e_{l})-{\cal W}_{\rm eff}(e_{k})\right] \nonumber \\
& = & \frac{1}{2\pi}\left[N(e_{l}-e_{k})-\sum_{i=1}^{N}m_{i}\log\left(\frac{e_{l}+m_{i}}
{e_{k}+m_{i}}\right)\right]
\label{zkl}
\end{eqnarray} 
At least for $|m_{i}-m_{j}|<<\Lambda$, 
the BPS property ensures that the soliton states 
of the $CP^{N-1}$ $\sigma$-model persist in the massive case. 
Typically the only effect of introducing small non-zero values of $|m_{i}-m_{j}|$ 
is to introduce splittings between the 
degenerate states which form multiplets of $SU(N)$ in the massless limit. 
This reflects the fact that the twisted masses explicitly break the $SU(N)$ global symmetry 
down to its maximal abelian subgroup. More generally, the global description 
of the BPS spectrum  may be complicated by the presence of 
curves of marginal stability in the parameter space on which the number of BPS states can change
\paragraph{}
Even for small twisted masses, the 
branch-cuts in the logarithms appearing in (\ref{zkl}) lead to an ambiguity in the BPS 
spectrum. In particular the ambiguity in the central charge is equal to 
$i\sum_{N=1}^{\infty}m_{i}n_{i}$ where the choice of integers $n_{i}$ corresponds to 
a choice of branch for each of the $N$ logarithms in (\ref{zkl}). As explained 
by Hanany and Hori in \cite{hnh}, this ambiguity signals the fact that solitons can 
carry integer values of the global $U(1)$ charges $S_{i}$ in addition to 
their topological charges. Clearly this is related to the existence of BPS dyons 
at weak-coupling discussed in the previous sections. As explained above, the BPS masses 
are independent of the gauge coupling $e$ and therefore (\ref{zkl}) should also provide an 
exact description of the mass spectrum even in the $\sigma$-model 
limit $e>>\Lambda$ In particular, the modified formula for the central charge,   
\begin{equation}
Z=2\Delta{\cal W}+\sum_{i=1}^{N}m_{i}S_{i}
\label{ccharge3}
\end{equation}
should apply for all values of the parameters. Indeed, this 
has exactly the same form as the formula (\ref{ccharge}) for the central 
charge of the classical theory considered in Section 3; the two formulae differ 
only by the replacement of the classical twisted superpotential (\ref{tfterm}) 
by its quantum counterpart (\ref{weff2}). In terms of the Noether and topological charges,  
$\vec{S}$ and $\vec{T}$, introduced in Section 3, (\ref{ccharge}) becomes, 
\begin{equation}
Z=-i(\vec{m}\cdot\vec{S}+\vec{m}_{D}\cdot\vec{T})
\label{ccharge5}
\end{equation}
with, 
\begin{equation}
\vec{m}_{D}=2i\left({\cal W}_{\rm eff}(e_{1}),{\cal W}_{\rm eff}(e_{2}),
\ldots {\cal W}_{\rm eff}(e_{N})\right)
\label{set}
\end{equation}         
\paragraph{}
The exact mass formula for all BPS states in the theory is $M=|Z|$ with 
$Z$ given by equations (\ref{ccharge5}) and (\ref{set}). In particular, for 
$|m_{i}-m_{j}|>>\Lambda$, this formula can be directly 
compared with the results of the semiclassical analysis given in the previous sections.   
As before, it is helpful to consider an explicit example for the case $N=2$.
Setting $m_{2}=-m_{1}=m/2$ the vacuum equation (\ref{vacuumeq}) becomes 
\begin{equation}
\sigma^{2}-\frac{m^{2}}{4}= \tilde{\Lambda}^{2}
\label{ven=2}
\end{equation}
Thus the two supersymmetric vacua are located at $\sigma=
\pm \sqrt{m^{2}/4+\tilde{\Lambda}^{2}}$. 
For $m=0$, this reduces to $\sigma=\pm \tilde{\Lambda}$ which is a special case of  
(\ref{sigma1}). On the other had, for $|m|>>\Lambda$ the two vacua are approximately located at 
$\sigma=\pm m/2$ which agrees with the classical analysis of Section 3.  The central 
charge  is $Z=-i(mS+m_{D}T)$ where $S=(S_{1}-S_{2})/2$, $T=(T_{1}-T_{2})/2$ and
\begin{equation}
m_{D}=-\frac{i}{\pi}\left[ \sqrt{m^{2}+4\tilde{\Lambda}^{2}} +\frac{m}{2}
\log\left(\frac{m- \sqrt{m^{2}+4\tilde{\Lambda}^{2}}}{m+  \sqrt{m^{2}+4\tilde{\Lambda}^{2}}}
\right)\right]
\label{mpm}
\end{equation}
\paragraph{}
In the weak-coupling limit, $m>>\tilde{\Lambda}$, this yields the expansion,
\begin{equation}
m_{D}=\frac{im}{\pi}\left[\log\left(\frac{m}{\tilde{\Lambda}}\right)+i\pi 
+\sum_{k=1}^{\infty}\,c_{k}\left(\frac{\tilde{\Lambda}}{m}\right)^{2k}\right]
\label{infserB}
\end{equation}
with $c_{k}=(-1)^{k}(2k-2)!/(k!)^{2}$. The first term in (\ref{infserB}) 
agrees with the one-loop formula 
for $m_{D}$ derived in Section 5 and, in principle, predicts the value 
of the undetermined additive constant $C$ appearing in the one-loop 
soliton mass formula (\ref{finalmass}).  
The remaining series of terms have exactly the right 
form to be interpreted as the leading semiclassical contributions of $\sigma$-model instantons. 
The explicit result for the coefficients $c_{k}$ constitutes an infinite set of predictions 
which could be tested against first-principles instanton calculations.     
The semiclassical analysis of the previous sections established that the 
weakly-coupled theory at large $|m|$ contains an infinite tower of dyon states with 
charge $(S,T)$ where $T=\pm 1$ and $S$ is an integer. 
As $m$ describes a large circle in the complex 
plane the dyon spectrum under goes a non-trivial monodromy $(S,T)\rightarrow (S-2T,T)$. 
The BPS spectrum also 
exhibits another phenomenon which is familiar in the context of ${\cal N}=2$ theories in 
four dimensions: strongly coupled vacua with a massless dyon. Equation (\ref{mpm}) 
shows that a dyon state becomes massless at $m=\pm 2\tilde{\Lambda}$. As in ${\cal N}=2$ 
SUSY Yang-Mills \cite{SW1}, 
each dyon can be made massless at one of these two points by successively 
applying the weak-coupling monodromy. 
We can choose conventions so that a dyon with even $S$ can become massless at 
$m=+2\tilde{\Lambda}$ while one with odd $S$ becomes massless $m=-2\tilde{\Lambda}$.
\paragraph{}
In addition to the dyon states, the weak coupling theory also has an 
elementary particle with $(S,T)=(1,0)$. According to the above analysis, 
The mass of this state is equal to its tree level value $|m|$. Naively, this 
implies that the theory contains a massless particle for $m=0$. However, for 
$m=0$ we know that the theory reduces to the supersymmetric $CP^{1}$ 
$\sigma$-model which certainly has a mass gap. In fact, 
for $|m|<<\Lambda$, the BPS spectrum should reduce to that of the 
$\sigma$-model which has only two stable BPS states as opposed to the infinite 
number of states present at weak coupling.   
As usual this discrepancy can be resolved if the two regions of parameter space 
are separated by a curve of marginal stability (CMS) on which the 
number of BPS states can change.  In the following, at least for the case of the 
elementary particle, we will perform an explicit check that the required CMS exists. We will 
consider  paths in the complex $m$-plane which connect the weak and strong coupling regions 
and show that there is at least one point on each path at which the elementary particle is 
exactly at threshold to decay. The paths in question are radial lines parametrized by $r>0$ 
with $m/2\tilde{\Lambda}=r\exp(i\delta)$ for a fixed value of 
$\delta\in [0,2\pi]$. Rather than performing a complete check, 
we will restrict our attention to the simplest 
cases $\delta=0$,$\pi$ and $\delta=\pi/2$, $3\pi/2$.   
At the end of this section, a more general argument  
for the existence of the required CMS will be given which 
exploits the connection between the BPS spectra considered here and 
those of ${\cal N}=2$ theories in four dimensions. 
\paragraph{}
For $\delta=0$, the central charge determined by (\ref{mpm}) becomes 
\begin{equation}
Z=-i\left(m(S+\frac{1}{2}T)-\frac{2i}{\pi}\tilde{\Lambda}
f\left(\frac{m}{2\tilde{\Lambda}}\right)T\right)
\label{zz}
\end{equation}
where,   
\begin{equation}
f(r)=\sqrt{1+r^{2}} +\frac{r}{2}
\log\left(\frac{\sqrt{1+r^{2}}-r}
{\sqrt{1+r^{2}}+r}
\right)
\label{mpm2}
\end{equation}  
is a continuous, real function of $r$ for $r>0$. 
The masses of states with $(S,T)$ charges $(1,0)$, 
$(0,1)$ and $(1,-1)$ obey the inequality,
\begin{equation} 
M_{(1,0)}=m \leq 2\sqrt{\left(\frac{m}{2}\right)^{2}+\frac{4\tilde{\Lambda}^{2}}{\pi^{2}}
f^{2}\left(\frac{m}{2\tilde{\Lambda}}\right)}
=M_{(0,1)}+
M_{(1,-1)}
\label{ineq}
\end{equation}
with equality if and only if $f=0$. Thus the elementary particle with charge $(1,0)$ is stable 
against decay into the dyon/soliton states with charges $(0,1)$ and $(1,-1)$ as long 
as $f\neq 0$. However, for $f=0$ the elementary particle is exactly at the threshold for 
decay into these products. Hence, to prove that this path in parameter space passes through 
a CMS we must find a zero of $f(r)$ for some positive $r$. The 
asymptotic behaviour of $f(r)$ for large positive $r$ is $f\sim -r\log r$ which is arbitrarily 
large and negative. On the other hand we have $f(0)=1>0$. As $f$ is continuous, 
the existence of the required zero follows immediately from the intermediate value theorem. 
A trivial modification of this argument shows that a similar point exists on the $\delta=\pi$ 
path. For $\delta=\pi/2$ we have $m/2\tilde{\Lambda}=i\nu$ for 
real positive $\nu$. In this case, $m_{D}=0$ and a dyon 
becomes massless at $\nu=1$ and the elementary particle 
is at threshold to decay into the same products as the $\delta=0$ case. 
The other singular point lies on the path with $\delta=3\pi/2$ and a similar argument 
applies.        
\paragraph{}
In Section 4, we found a classical correspondence between the BPS states of the 
two-dimensional theory with $N$ chiral multiplets and those of four-dimensional 
${\cal N}=2$ SYM with gauge group $SU(N)$. In the following we will investigate 
whether this correspondence persists at the quantum level. The exact BPS spectrum 
of a quantum ${\cal N}=2$ theory in four dimensions is determined by the periods 
of an elliptic curve. We would therefore like to interpret the exact central charges 
(\ref{zkl}) of the two-dimensional theory as periods of a complex curve and determine 
which, if any,  four-dimensional theory is described by the same curve. 
Before presenting a curve with the required property, we briefly review our expectations 
of the corresponding four-dimensional theory.   
As explained at the end of the previous section, the naive extension of 
the classical correspondence found above is incorrect because the 
R-symmetry anomalies of the two theories in question are different.            
In order to match the pattern of R-symmetry breaking of the two-dimensional theory it 
is necessary to go to a four-dimensional ${\cal N}=2$ theory with gauge group 
$SU(N)$ and $N$ additional hypermultiplets in the fundamental representation. 
To specify the correspondence we should determine the relation between the parameters (or moduli 
in the four-dimensional case) of the two theories.     
In the classical case, we identified the $N$  complex eigenvalues, $a_{i}$, 
of the adjoint scalar VEV in the four-dimensional theory with the $N$ twisted masses, $m_{i}$, 
of the two-dimensional theory. 
However, the story is more complicated for the proposed quantum correspondence
as the four-dimensional theory with $N$ hypermultiplets also has $N$ additional parameters: 
the hypermultiplet masses $\mu_{i}$. Clearly some additional conditions on the parameters/moduli 
of the four-dimensional theory are required to specify the correspondence. 
\paragraph{}
In fact it is straightforward to find a complex curve whose non-vanishing periods are 
given by (\ref{zkl}). For 
each $N$, the curve is defined by a polynomial equation of order $2N$ in $x$  
\begin{equation}
y^{2}=\frac{1}{4}\left[\prod_{i=1}^{N}(x+m_{i})-\tilde{\Lambda}^{N}\right]^{2}=
\prod_{i=1}^{N}(x-e_{i})^{2}
\label{curveb}
\end{equation}   
The roots, $e_{i}$ are those defined in (\ref{vacuumeq}) above with the replacement 
$\sigma \rightarrow x$.   
As expected, this curve describes and $SU(N)$ 
${\cal N}=2$ SQCD in four-dimensions with $N$ hypermultiplets having masses 
$\mu_{i}=-m_{i}$. The four-dimensional 
theory is at a special point on its Coulomb branch where 
a maximal number of one-cycles vanish and $N$ massless hypermultiplets
\footnote{Note however that the BPS states of the two-dimensional theory correspond 
to the non-vanishing cycles of the curve and are all massive as expected.} 
appear.  This point is the root of the baryonic Higgs branch \cite{APS}. 
Correspondingly, the polynomial (\ref{curveb}) 
is a perfect square; each of the $N$ distinct roots $e_{i}$ occurs exactly twice.  
By performing the change of variables, 
\begin{equation}
t=y-\frac{1}{2}\left[\prod_{i=1}^{N}(x+m_{i})+\Lambda^{N}\right]
\label{tt}
\end{equation}
the curve can be rewritten in an alternative form which appears naturally in Witten's 
M-theory construction of the model \cite{W4} and also in the context of the relation 
between ${\cal N}=2$ theories and integrable systems \cite{INT}, 
\begin{equation}
(t-\tilde{\Lambda}^{N})\left(t-\prod_{i=1}^{N}(x+m_{i})\right)=0
\label{curvec}
\end{equation}
The non-vanishing periods can then be written as integrals over the Seiberg-Witten 
differential. In terms of the variables $x$ and $t$, the periods are,  
\begin{equation}
{\cal P}_{kl}=\int_{e_{k}}^{e_{l}}d\lambda_{\rm SW}=
\int_{e_{k}}^{e_{l}} x \frac{dt}{t}
\label{swdiff}
\end{equation}
which gives 
\begin{eqnarray}
{\cal P}_{kl} & = & \sum_{i=1}^{N}\,\int_{e_{k}}^{e_{l}} \frac{x}{x+m_{i}} 
\nonumber \\
& = & N(e_{l}-e_{k})-\sum_{i=1}^{N}m_{i}\log\left(\frac{e_{l}+m_{i}}
{e_{k}+m_{i}}\right)
\label{fr}
\end{eqnarray}
Thus, up to an overall numerical factor, 
the period ${\cal P}_{kl}$ is equal to the central charge $Z_{kl}$ given in (\ref{zkl}) 
as advertised above.
\paragraph{}
Several features of this correspondence require further comment.  
First, the twisted masses of the two-dimensional theory are now identified as 
hypermultiplet masses $\mu_{i}$ in the four-dimensional theory. 
Even in the weak-coupling limit, this seems to disagree with 
the classical correspondence, where the twisted masses were identified instead with the 
adjoint scalar VEVs $a_{i}$. In fact, there is actually no disagreement because, at 
least in the weak coupling regime, we have $a_{i}=\mu_{i}$ at the singular point. Second, as in 
the classical case, only the massive BPS states of the four-dimensional theory have 
counterparts in the two dimensional theory. 
For $|\mu_{i}-\mu_{j}|>>\Lambda$, the curve (\ref{curveb}) describes a 
weakly-coupled four-dimensional theory with $N$ massless quarks 
and an infinite number of stable BPS states. For $|\mu_{i}-\mu_{j}|<<\Lambda$ 
the curve describes a strongly-coupled theory in four-dimensions with 
$N$ massless dyons. At least in the 
simplest case $N=2$, which corresponds to gauge group $SU(2)$ in four dimensions, it is known 
that the strongly-coupled theory has only a finite number of stable BPS states \cite{SW2,BF2,BF}. 
This is possible because 
the two regions of parameter/moduli space in the four-dimensional theory are disconnected 
by a surface of marginal stability of real codimension one. The explicit form of this surface has 
recently been determined in \cite{BF}. In the above, we found that a curve of marginal 
stability in the complex mass plane was required in order to produce a consistent description 
of the BPS spectrum of the two-dimensional theory with $N=2$. In this case, the existence 
of the required CMS can be deduced from the analysis of the corresponding 
four-dimensional theory given in \cite{BF}.
\paragraph{}
The author acknowledges helpful discussions with Amihay Hanany, Dave Tong and 
Stefan Vandoren.    
\section*{Appendix A}
\paragraph{}
In Section 2, we considered the most general supersymmetric action with up to two 
derivatives or four-fermions for chiral superfields $\Phi_{i}$,   
\begin{equation}
{\cal L}={\cal L}_{D}+{\cal L}_{F}=\int\,d^{4}\theta\, K(\Phi_{i},\bar{\Phi}_{i}) \,+\, 
\int\,d^{2}\theta\, W(\Phi_{i}) \, \, + \int\,
d^{2}\bar{\theta}\, \bar{W}(\bar{\Phi}_{i})
\label{aterm}
\end{equation}
Minimal coupling of the 
each chiral multiplet to an abelian gauge superfield $V$ (with equal charges) 
is generated by the replacement.  
\begin{equation}
\Phi_{i} \rightarrow \exp(V)\Phi_{i}
\label{amincoupling}
\end{equation}
This Appendix provides a formula for the resulting 
Lagrangian in terms of component fields. 
In particular, explicit results will be given for 
the component Lagrangian of the $CP^{1}$ $\sigma$-model 
with twisted mass terms. 
\paragraph{}   
As in Section 1, we will derive the two-dimensional component 
Lagrangian by dimensional reduction from a theory in four-dimensions with ${\cal N}=1$ 
supersymmetry. 
The component expansion of the {\em four-dimensional} Lagrangian defined by 
${\cal L}_{D}$ with the minimal coupling prescription (\ref{amincoupling}) 
can be read off from formula (24) of \cite{BW}. 
Rewriting this formulae in the notation of Section 1 yields, 
\begin{eqnarray}
{\cal L}_{D} &= & g_{i\bar{j}}\left(-{ D}_{m}\phi^{i}{ D}^{m}\bar{\phi}^{j}-
\bar{\psi}^{j}_{\dot{\alpha}}\bar{\sigma}^{m\dot{\alpha}\alpha}{\cal D}_{m}\psi^{i}_{\alpha} 
+F^{i}\bar{F}^{j}\right) \nonumber \\ 
& & -\frac{D}{2}\left(\phi^{i}\frac{\partial K}{\partial \phi^{i}}+\bar{\phi}_{i}\frac{\partial K}
{\partial\bar{\phi}_{i}}\right)+
i g_{i\bar{j}}\left(\bar{\phi}^{j}\psi^{i\alpha}\chi_{\alpha}+
\phi^{i}\bar{\psi}^{j}_{\dot{\alpha}}
\bar{\chi}^{\dot{\alpha}}\right) \nonumber \\
& & -\frac{1}{2}F^{i}g_{i\bar{l}}\Gamma^{\bar{l}}_{\bar{j}\bar{k}}
\bar{\psi}^{j}_{\dot{\alpha}}\bar{\psi}^{k\dot{\alpha}}  
-\frac{1}{2}\bar{F}^{i}g_{l\bar{i}}\Gamma^{l}_{jk}{\psi}^{j\alpha}{\psi}^{k}_{\alpha}
+\frac{1}{4}g_{i\bar{j},k\bar{l}}{\psi}^{i\alpha}{\psi}^{k}_{\alpha}
\bar{\psi}^{j}_{\dot{\alpha}}\bar{\psi}^{l\dot{\alpha}} 
\label{long}
\end{eqnarray}
with the usual definitions for the tensors which arise naturally on a \Ka\ manifold.  
\begin{eqnarray}
g_{i\bar{j}} & = & \frac{\partial^{2}  K}{\partial \phi^{i}\partial\bar{\phi}^{j}} \nonumber \\
g_{i\bar{j},k} & = &\frac{\partial}{\partial\phi^{k}}\,g_{i\bar{j}}
=g_{l\bar{j}}\Gamma^{l}_{ik}\nonumber \\
g_{i\bar{j},\bar{k}} & = &\frac{\partial}{\partial\bar{\phi}^{k}}\,g_{i\bar{j}} =g_{i\bar{l}}
\Gamma^{\bar{l}}_{\bar{j}\bar{k}}\nonumber \\
g_{i\bar{j},k\bar{l}} & = &\frac{\partial^{2}}{\partial \phi^{k}\partial\bar{\phi}^{l}} 
\, g_{i\bar{j}}
\label{long2}
\end{eqnarray}
Here $D_{m}$ denotes the gauge-covariant derivative for a $U(1)$ gauge field 
$V_{m}$: $D_{m}=\partial_{m}+iV_{m}$. 
The fermion kinetic term in (\ref{long}) contains a derivative which is also 
covariant with respect to general coordinate transformations, 
\begin{equation}
{\cal D}_{m}\psi^{i}_{\alpha}=D_{m}\psi^{i}_{\alpha}+
\Gamma^{i}_{jk}(D_{m}\phi^{j})\psi^{k}_{\alpha}
\label{acov}
\end{equation}
The F-term Lagrangian is not modified by minimal coupling and is given by, 
\begin{eqnarray}
{\cal L}_{F} & = & F^{i}\frac{\partial W}{\partial \phi^{i}} + 
\bar{F}^{i}\frac{\partial \bar{W}}{\partial \bar{\phi}^{i}}
-\frac{1}{2}\frac{\partial^{2} W}{\partial \phi^{i}
\partial\phi^{j}}{\psi}^{i\alpha}{\psi}^{j}_{\alpha}
-\frac{1}{2}\frac{\partial^{2}\bar{W}}{\partial \bar{\phi}^{i}\partial\bar{\phi}^{j}}
\bar{\psi}^{i}_{\dot{\alpha}}\bar{\psi}^{j\dot{\alpha}}
\label{long3}
\end{eqnarray}
\paragraph{}
As in Section 1, the two-dimensional fields in each chiral multiplet are a complex scalar 
$\phi^{i}$, a Dirac fermion with components 
$\psi^{i}_{-}=\psi^{i}_{1}$ and $\psi^{i}_{+}=\psi^{i}_{2}$ as well as a complex auxiliary field 
$F^{i}$. The gauge multiplet fields include the complex scalar $\sigma=V_{1}-iV_{2}$, 
and the 2D gauge field with components $v_{0}=V_{0}$ and $v_{1}=V_{3}$ as well as a 
fermion with components $\chi_{\alpha}$ 
and a real auxiliary field $D$. The two-dimensional Lagrangian 
is then obtained from (\ref{long}) by dimensional reduction in the 
$X_{1}$ and $X_{2}$ directions.  
\paragraph{}  
In Sections 3, 4 and 5, the case of a $U(1)$ gauge theory with two chiral multiplets is 
analysed in detail. In this case the low-energy effective theory is a variant of the 
supersymmetric $CP^{1}$ $\sigma$-model which includes twisted mass terms. The Lagrangian 
is a D-term for a single chiral superfield\footnote{Hopefully denoting the chiral 
superfield with the same letter as the superpotential will not cause confusion as the latter is 
zero in this example.} $W$ with components 
$(w,\psi_{\alpha},F)$ and minimal coupling to a background 
gauge superfield $\hat{V}$ with expectation value, 
\begin{equation} 
\langle \hat{V} \rangle = -\theta^{\alpha}\sigma^{m}_{\alpha\dot{\alpha}}
\bar{\theta}^{\dot{\alpha}}\hat{V}_{m}
\label{abackground}
\end{equation}
where $\hat{V}_{1}={\rm Re}(m)$, $\hat{V}_{2}=-{\rm Im}(m)$ and 
$\hat{V}_{0}=\hat{V}_{3}=0$. Note that the fermion and auxiliary field components of 
$\hat{V}$ are zero. The superspace Lagrangian is obtained by setting $N=2$ in (\ref{d4}), 
\begin{equation}
{\cal L}=r \int\,d^{4}\theta\, 
\log\left(1+\bar{W}
\exp \left(2\langle\hat{V}\rangle\right) W\right)
\label{ad4}
\end{equation}   
The component Lagrangian follows from Equations 
(\ref{long}) and (\ref{long2}) with $\Phi_{1}=W$ $V=\langle \hat V \rangle$, 
$K(W,\bar{W})=r\log(1+\bar{W}{W})$. The final result is,  
\begin{eqnarray}
{\cal L} &= & {\cal L}^{(0)}+ {\cal L}^{(2)}+{\cal L}^{(4)} \nonumber \\ 
{\cal L}^{(0)} & = & 
-\frac{1}{\rho^{2}}\left[r\left(\partial_{\mu}\bar{w}\partial^{\mu}w 
+|m|^{2}|w|^{2}\right)+\frac{\theta}{2\pi}\varepsilon^{\mu\nu} 
\partial_{\mu}\bar{w}\partial_{\nu}w\right] \nonumber \\
{\cal L}^{(2)} & = &\, \frac{1}{\rho^{2}}\left[ 
i\bar{\psi}_{-}
\left(\partial_{+}-\frac{2i}{\rho}{\rm Im}(\bar{w}\partial_{+}w)\right)\psi_{-}
+i\bar{\psi}_{+}
\left(\partial_{-}-\frac{2i}{\rho}{\rm Im}(\bar{w}\partial_{-}w)\right)\psi_{+}\right] 
\nonumber \\
& & \qquad{} \qquad{} \qquad{} \qquad{} -\frac{i}{\rho^{2}}  
\left(1-\frac{2|w|^{2}}{\rho}\right) 
\left(m\bar{\psi}_{-}\psi_{+}- \bar{m}\bar{\psi}_{-}\psi_{+}\right) \nonumber \\
{\cal L}^{(4)} & =&\, \frac{r}{\rho^{2}}\left[\left(
F-\frac{2\bar{w}\psi_{-}\psi_{+}}{\rho}\right)\left(
\bar{F}-\frac{2w\bar{\psi}_{-}\bar{\psi}_{+}}{\rho}\right) 
+\frac{2}{\rho^{2}}\psi_{-}\psi_{+}\bar{\psi}_{-}\bar{\psi}_{+}\right] \nonumber \\
\label{big}
\end{eqnarray}
where $\rho=1+|w|^{2}$ and $\partial_{\pm}=\partial_{0}\pm\partial_{1}$. 
\section*{Appendix B}
\paragraph{}
This Appendix provides further details of the small-fluctuation expansion 
around the soliton and dyon background. Specifically, we will start from the 
component Lagrangian (\ref{big}) derived in the previous appendix and expand the fields 
$w$, $\psi_{\alpha}$ and $F$ around their values in the general time-dependent 
dyon solution,  
\begin{equation}
w_{\rm cl}= \exp\left(\sqrt{|m|^{2}-\omega^{2}}\,x \,+\,i\omega t\right)
\label{bwcl}
\end{equation}
and $\psi^{\alpha}_{\rm cl}=F_{\rm cl}=0$. In this way we will derive the formulae 
(\ref{expand}-\ref{delb}) 
which were used in the calculation of quantum corrections to 
the soliton mass in Section 5. In the process we will also derive (\ref{fquadratic}) 
which was the starting point for the calculation of the 
weak-coupling monodromy of the dyon spectrum 
given in the text. We consider the expansion of the bose and 
fermi fields in turn.  
\subsubsection*{Bosons}
\paragraph{} 
In Section 4, the bosonic Lagrangian ${\cal L}^{(0)}$ is simplified by the change of variables, 
\begin{equation}
w=\tan\left(\frac{\varphi}{2}\right)\exp(i\alpha)
\label{bcov}
\end{equation}
The Lagrangian is given in terms of $\varphi$ and $\alpha$ in Equation (\ref{sg}) 
of Section 4. The classical values of these fields are $\varphi_{\rm cl}=\varphi_{\omega}(x)$ 
and $\alpha_{\rm cl}=\omega t$ where $\varphi_{\omega}(x)$ is defined in Equation (\ref{risky}).  
To determine the Lagrangian for the 
quadratic fluctuations of the bosonic fields, it is useful to perform 
yet another change of variables and write the Lagrangian in terms of 
a three component unit vector $\hat{\bf{n}}=(n_{1},n_{2},n_{3})$, 
$\hat{\bf n}\cdot\hat{\bf n}=1$. The scalar field $w$ is given in terms of 
the components of $\hat{\bf n}$ by, 
\begin{equation}
w=\frac{n_{1}+in_{2}}{1-n_{3}}
\label{wn}
\end{equation}
In addition we may express ${\bf n}$ in terms of $\varphi$ and $\alpha$ as, 
\begin{equation}
{\bf n}=(\sin\varphi\cos\alpha ,\,\sin\varphi\sin\alpha,\,-\cos\varphi)
\label{bnphi}
\end{equation}
With this change of variables, (\ref{bosonic}) becomes (for $\theta=0$) 
\begin{equation} 
{\cal L}^{(0)}= -\frac{r}{4}\left[\partial_{\mu}\hat{\bf n}\cdot
\partial^{\mu}\hat{\bf n}+|m|^{2}(1-n^{2}_{3})\right]
\label{o3}
\end{equation} 
For $m=0$, this is the action of the $O(3)$ $\sigma$-model 
and the change of variables reflects the standard equivalence between 
$\sigma$-models with target space $CP^{1}$ and $O(3)$. 
The effect of a non-zero twisted mass $m$ is to introduce a potential on the 
target manifold which has minima at the north and south poles 
$n_{3}=\pm 1$. These points are the two supersymmetric vacua of the theory. 
\paragraph{}
Next we expand ${\bf n}$ around its classical value, 
\begin{equation}
{\bf n}_{\rm cl}= 
(\sin(\varphi_{\omega})\cos(\omega t),
\sin(\varphi_{\omega})\sin(\omega t), 
-\cos(\varphi_{\omega})) 
\label{bncl}
\end{equation}
It is convenient to parametrize the fluctuations of the constrained field ${\bf n}$ 
in terms of two unconstrained real variables $u_{1}$ and $u_{2}$,  
\begin{eqnarray}
{\bf n} &= & {\bf n}_{\rm cl}+ 
\delta{\bf n} \nonumber \\ & = & 
(\sin(\varphi_{\omega}+u_{2})\cos(\omega t)-u_{1}\sin(\omega t),
\sin(\varphi_{\omega}+u_{2})\sin(\omega t) +u_{1}\cos(\omega t),-
\cos(\varphi_{\omega}+u_{2})) \nonumber \\
\label{delnphi}
\end{eqnarray}
 Expanding to quadratic order in bosonic 
fluctuations we have,  
\begin{equation}
{\cal L}^{(0)}={\cal L}^{(0)}[\varphi_{\rm cl},\alpha_{\rm cl}]- \frac{r}{4}
\vec{u}^{\rm T}\cdot {\cal M}_{\rm B}\cdot \vec{u}\, + \, O(\vec{u}^{3})
\end{equation}
where $\vec{u}^{\rm T}=(u_{1},u_{2})$ and 
\begin{equation}
{\cal M}_{\rm B}=\left(\begin{array}{cc} \frac{\partial^{2}}{\partial t^{2}}
+\Delta_{\rm B}(\omega) & 2\omega\cos(\varphi_{\omega})\frac{\partial}{\partial t} \\
-2\omega\cos(\varphi_{\omega})\frac{\partial}{\partial t} &  
\frac{\partial^{2}}{\partial t^{2}}+\Delta_{\rm B}(\omega)\end{array}\right) 
\label{bmb}
\end{equation}
with, 
\begin{equation}
\Delta_{\rm B}(\omega)=-\frac{\partial^{2}}{\partial x^{2}}+(|m|^{2}-\omega^{2})
\cos(2\varphi_{\omega})
\label{bdelb}
\end{equation}
Setting $\omega=0$, (\ref{bmb}) reproduces the bosonic terms in (\ref{expand}-\ref{delb})
\subsubsection*{Fermions}
\paragraph{}
The terms in the expansion which are bilinear in the fermion fields $\psi_{\alpha}$ 
and $\bar{\psi}_{\dot{\alpha}}$ and independent of the bosonic fluctuations may be 
calculated by replacing $w$ by its classical value $w_{\rm cl}$ in the fermion bilinear Lagrangian 
${\cal L}^{(2)}$ which appears in (\ref{big}), 
\begin{eqnarray}
{\cal L}^{(2)}[\psi,\bar{\psi},w_{cl}] &  =  &\frac{1}{\rho_{\rm cl}^{2}}\left[ 
i\bar{\psi}_{-}
\left(\partial_{+}-\frac{2i}{\rho_{\rm cl}}{\rm Im}(\bar{w}_{\rm cl}\partial_{+}
w_{\rm cl})\right)\psi_{-}
+i\bar{\psi}_{+}
\left(\partial_{-}-\frac{2i}{\rho_{\rm cl}}{\rm Im}(\bar{w}_{\rm cl}\partial_{-}w_{\rm cl}
)\right)\psi_{+}\right]  \nonumber \\ 
& & \qquad{} \qquad{} \qquad{} \qquad{} -\frac{i}{\rho_{\rm cl}^{2}}  
\left(1-\frac{2|w_{\rm cl}|^{2}}{\rho_{\rm cl}}\right) 
\left(m\bar{\psi}_{-}\psi_{+}-\bar{m}\bar{\psi}_{-}\psi_{+}\right) 
\label{bbig2}
\end{eqnarray}
where $\rho_{\rm cl}=1+|w_{\rm cl}|^{2}$. Introducing the $\gamma$-matrices 
$\gamma^{0}=i\sigma_{2}$, $\gamma^{1}=-\sigma_{1}$ and $\gamma^{5}=\gamma^{0}\gamma^{1}
=-\sigma_{3}$ 
and Dirac fermions; 
\begin{eqnarray}
\tilde{\Psi}=\left(\begin{array}{c} \psi_{-} \\ \psi_{+} \end{array}\right) & \qquad{} \qquad{} 
\qquad{} & \tilde{\Psi}^{\dagger}=\left(\begin{array}{c} \bar{\psi}_{-} \\ \bar{\psi}_{+} 
\end{array}\right)
\label{bdir}
\end{eqnarray}
Defining $\Psi=\exp(i\pi\gamma^{5}/4)\tilde{\Psi}$ and $\bar{\Psi}=\Psi^{\dagger}\gamma^{0}$ we get 
\begin{equation}
{\cal L}^{(2)}[\psi,\bar{\psi},w_{\rm cl}]=
-\frac{ir}{4}\bar{\Psi}_{\alpha}
\hat{\cal M}_{\rm F}^{\alpha\beta}\Psi_{\beta}
\label{bexpand2}
\end{equation}  
with 
\begin{equation}
\hat{\cal M}_{\rm F} = \gamma^{\mu}
\left(\partial_{\mu}+\frac{2i}{\rho_{\rm cl}}{\rm Im}(\bar{w}_{\rm cl}\partial_{\mu}
w_{\rm cl})\right)+\left(1-\frac{2|w_{\rm cl}|^{2}}{\rho^{2}_{\rm cl}}\right)
\left[{\rm Re}(m)I+i{\rm Im}(m)\gamma^{5} \right]\Psi
\label{blf2}
\end{equation}
substituting for $w_{\rm cl}$ as in (\ref{bwcl}) yields,
\begin{equation}
\hat{\cal M}_{\rm F}  = 
\gamma^{\mu}\partial_{\mu}+\cos\left(\varphi_{\omega}(x)\right)
\left[i\omega\gamma^{0}+{\rm Re}(m)I+i{\rm Im}(m)\gamma^{5}\right] 
\label{bres1}
\end{equation}
This is Equation (\ref{fquadratic}) of Section 5. 
In the special case of the static soliton $\omega=0$, we may rewrite (\ref{blf2}) 
in a form which makes the supersymmetry 
between boson and fermion fluctuations manifest with the following three steps:
\paragraph{}
{\bf 1:} In 
the special case of the static soliton with $\omega=0$, the 2D chiral anomaly for the 
Dirac operator (\ref{blf2}), which is derived in Section 5, vanishes. 
In  this case only, we may absorb all dependence on the 
phase of $m$ by a chiral redefinition of the Dirac spinors, 
\begin{eqnarray}
\Psi\rightarrow \Psi'=\exp\left(i\frac{\beta}{2}\gamma^{5}\right)\Psi 
& \qquad \qquad{} &  \bar{\Psi}\rightarrow \bar{\Psi}'=\bar{\Psi}
\exp\left(i\frac{\beta}{2}\gamma^{5}\right)
\label{bpsiprime}
\end{eqnarray}
where $\beta={\rm arg}(m)$. 
\paragraph{}
{\bf 2:} It is also convenient to choose a new basis for the spinors $\Psi$ and 
$\bar{\Psi}$ in which the $\gamma$-matrices become  
$\gamma^{0}=i\sigma_{2}$, $\gamma^{1}=-\sigma_{3}$ and 
$\gamma^{5}=\gamma^{0}\gamma^{1}=\sigma_{1}$. This is acomplished by the following 
unitary tranformation 
\begin{eqnarray}
\Psi'\rightarrow \Psi''=
\exp\left(i\frac{\pi}{4}\sigma_{2}\right)\Psi'
& \qquad \qquad{} &  \Psi'^{\dagger}\rightarrow \Psi''^{\dagger}=\Psi'^{\dagger} 
\exp\left(-i\frac{\pi}{4}\sigma_{2}\right)
\label{bpsidp}
\end{eqnarray}
\paragraph{}
{\bf 3:} We define two real Majorana fermions $\rho_{1\alpha}$ and $\rho_{2\alpha}$, with 
$\rho^{\dagger}_{i}=\rho^{\rm T}_{i}$, which are related to the Dirac fermion $\Psi''$ by 
$\Psi''=\rho_{1}+i\rho_{2}$. 
\paragraph{}
Finally, the bilinear action for the Majorana fermions $\vec{\rho}^{\,{\rm T}}_{\alpha}=
(\rho_{1\alpha},\rho_{2\alpha})$  is,  
\begin{equation}
{\cal L}^{(2)}[\psi,\bar{\psi},w_{\rm cl}]=-\frac{r}{4}\vec{\rho}^{\,{\rm T}}_{\alpha}\cdot
{\cal M}_{\rm F}^{\alpha\beta}\cdot\vec{\rho}_{\beta}
\label{bexpandm}
\end{equation}  
where
\begin{equation}
{\cal M}^{ij}_{\rm F}=  
\delta^{ij}\left(\begin{array}{cc} \frac{\partial}{\partial t} & D \\ -D^{T} &
\frac{\partial}{\partial t}\end{array}\right) 
\end{equation}
and 
\begin{eqnarray}
D=\frac{\partial}{\partial x}+|m|\cos\left(\varphi_{S}(x)\right) & \qquad{} & 
D^{T}=-\frac{\partial}{\partial x}+|m|\cos\left(\varphi_{S}(x)\right)
\label{bddt}
\end{eqnarray} 
This is gives the fermionic part of Equations (\ref{expand}-\ref{delb}). 

\end{document}